\documentclass[aps,prb,twocolumn]{revtex4-2}
\usepackage{amsmath}
\usepackage{amssymb}
\usepackage{graphicx}
\usepackage{dcolumn}
\usepackage{bm}
\usepackage[colorlinks=true,linkcolor=blue,anchorcolor=blue, citecolor=cyan,urlcolor=cyan]{hyperref}
\usepackage[mathlines]{lineno}
\usepackage{ulem}
\newcommand{\eq}[1]{\begin{equation} #1 \end{equation}}
\newcommand{\eqa}[1]{\begin{equation}\begin{aligned} #1 \end{aligned}\end{equation}}
\newcommand{\ua}{\uparrow}
\newcommand{\da}{\downarrow}
\newcommand{\bfk}{\boldsymbol{k}}
\newcommand{\cabs}[1]{\left|#1\right|}
\newcommand{\kb}[1]{\langle #1 \rangle}

\newcommand{\ks}[1]{|#1 \rangle}

\newcommand{\cs}[1]{\left(#1\right)}
\newcommand{\cm}[1]{\left[#1\right] }
\newcommand{\cl}[1]{\left\{#1\right\}}
\newcommand{\bsf}[1]{\boldsymbol{#1}}
\begin{document}
\title{Creation and Manipulation of Higher-Order Topological States by Altermagnets}
\author{Yu-Xuan Li}
\author{Yichen Liu}
\author{Cheng-Cheng Liu}
\email{ccliu@bit.edu.cn}
\affiliation{Centre for Quantum Physics, Key Laboratory of Advanced Optoelectronic Quantum Architecture and Measurement (MOE), School of Physics, Beijing Institute of Technology, Beijing 100081, China}
\begin{abstract}
We propose to implement tunable higher-order topological states in a heterojunction consisting of a two-dimensional (2D) topological insulator and the recently discovered altermagnets, whose unique spin-polarization in both real and reciprocal space and null magnetization are in contrast to conventional ferromagnets and antiferromagnets. Based on symmetry analysis and effective edge theory, we show that the special spin splitting in altermagnets with different symmetries, such as $d$-wave, can introduce Dirac mass terms with opposite signs on the adjacent boundaries of the topological insulator, resulting in the higher-order topological state with mass-domain bound corner states. Moreover, by adjusting the direction of the N\'{e}el vector, we can manipulate such topological corner states by moving their positions. By first-principles calculations, taking a 2D topological insulator bismuthene with a square lattice on an altermagnet MnF$_2$ as an example, we demonstrate the feasibility of creating and manipulating the higher-order topological states through altermagnets. Finally, we discuss the experimental implementation and detection of the tunable topological corner states, as well as the potential non-Abelian braiding of the Dirac corner fermions.
\end{abstract}
\maketitle
\textit{Introduction.---}Topological insulators (TIs), which have the helical edge states protected by time-reversal symmetry (TRS), set off an upsurge in topological matter research~\cite{hasan_colloquium_2010,qi_topological_2011}. Recently, the introduction of higher-order topological states has expanded the  topological matter research~\cite{sitte_topological_2012,zhang_surface_2013,benalcazar_quantized_2017,benalcazar_electric_2017,langbehn_reflection-symmetric_2017,song__2017,schindler_higher-order_2018,schindler_higher-order_2018-1,noh_topological_2018,hsu_majorana_2018,wang_high-temperature_2018,liu_majorana_2018,yan_majorana_2018,khalaf_higher-order_2018,ezawa_higher-order_2018,BJYang2018,xu_higher-order_2019,benalcazar_quantization_2019,pan_lattice-symmetry-assisted_2019,peng_floquet_2019,volpez_second-order_2019,zhang_helical_2019,zhu_tunable_2018,chen_universal_2020-1,wu_-plane_2020,zhang_mobius_2020,ren_engineering_2020,wu_boundary-obstructed_2020,nag_hierarchy_2021,yang_higher-order_2021,khalaf_boundary-obstructed_2021,zhang_stm_2023}. For the traditional first-order topological states like TIs, the difference between the dimensions of the topological boundary states and the bulk states is referred to as the codimension $d_c$ which satisfies $d_c=1$. In contrast, the higher-order topological states have a codimension $d_c$ greater than one. For example, a second-order topological insulator in $d$ dimensions exhibits the topologically protected hallmark boundary states of lower dimensionality $(d-2)$, such as corner states in two dimensions (2D) or hinge states in three dimensions (3D). Currently, only a few materials, such as SnTe~\cite{schindler_higher-order_2018-1}, bismuth~\cite{schindler_higher-order_2018}, EuIn$_2$As$_2$~\cite{xu_higher-order_2019} and MnBi$_{2n}$Te$_{3n+1}$~\cite{zhang_mobius_2020}, are predicted to be 3D higher-order topological insulators (HOTIs).  Experimental observation of the hinge states has so far been limited to bismuth~\cite{schindler_higher-order_2018,zhang_stm_2023}.  As for 2D HOTIs, various candidates have been proposed~\cite{qian_second-order_2021,Huang2022,Qian2022,Xie2021,TBG_hoti_2019,liu_higher-order_2021,sheng_two-dimensional_2019,ZJWang2022,XQChen2022}, including hydrogenated and halogenated 2D hexagonal group-IV materials~\cite{qian_second-order_2021,Huang2022}, Kekul\'{e}-ordered graphenes~\cite{qian_second-order_2021,mu_kekule_2022}, 2D transition metal dichalcogenides~\cite{Qian2022,Xie2021}, and twisted moir\'{e} superlattices ~\cite{TBG_hoti_2019,liu_higher-order_2021}, but the experimental confirmation of their corner states is still lacking.

Usual approaches employed to achieve HOTI states include introducing a Zeeman field into a first-order TI~\cite{sitte_topological_2012,zhu_tunable_2018,wu_-plane_2020,ren_engineering_2020} or harnessing the magnetic proximity effect to induce an exchange field within the TI~\cite{vobornik_magnetic_2011,zhang_surface_2013,liu_majorana_2018,chen_universal_2020-1}.  However, in the existing approaches, the manipulation of the topological corner states (TCSs) is a big challenge, which impedes their potential application such as in quantum information processing~\cite{kitaev_fault-tolerant_2003,Loss2013,wu_non-abelian_2020,wu_double-frequency_2020}.

\begin{figure}[]
    \centering
    \includegraphics[scale=0.4]{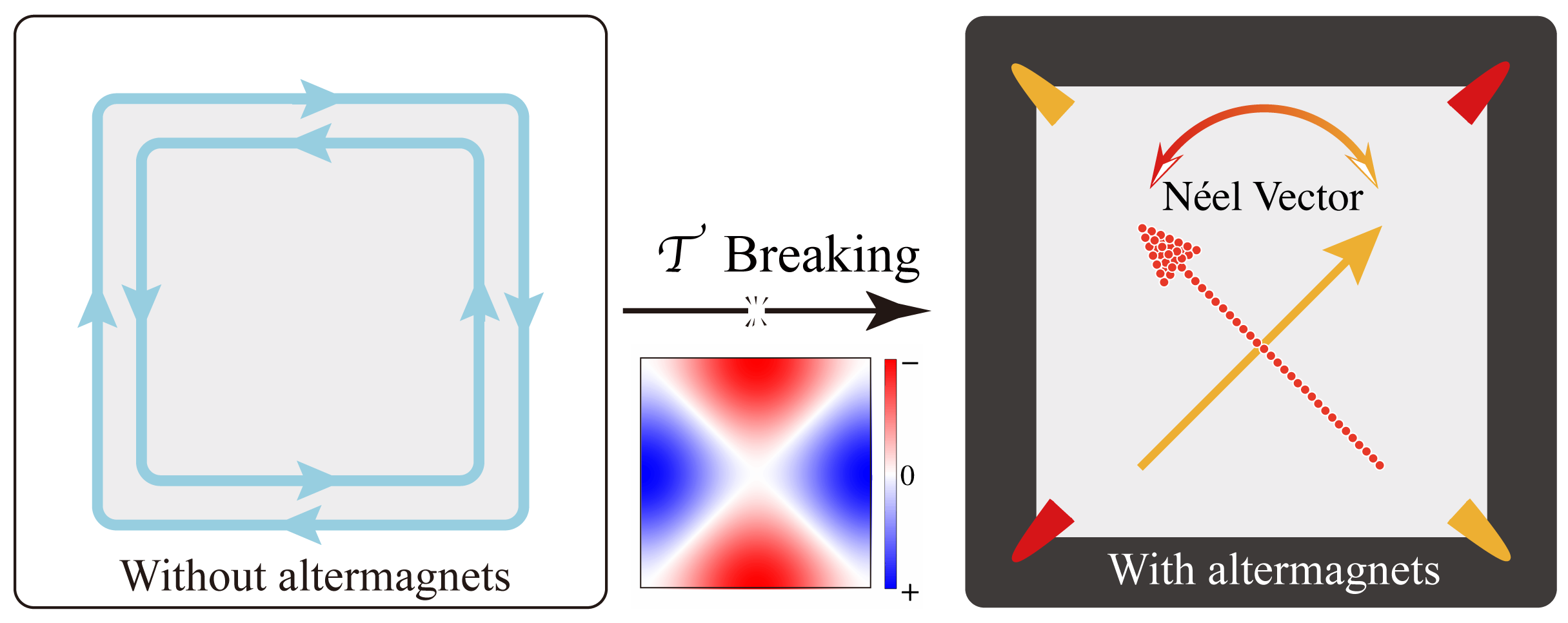}
    \caption{Left panel: A 2D topological insulator with the helical edge states protected by time-reversal symmetry. Middle panel: A proximitized altermagnet induces altermagnetism in the 2D topological insulator and breaks time-reversal symmetry. We take a $d$-wave altermagnet as an example. Right panel: When the N\'{e}el vector lies in the plane, the helical edge states open a gap with two in-gap states localized at the corners appearing, i.e. the topological corner states (in orange). The manipulation of these corner states (in red) can be achieved by adjusting the N\'{e}el vector around the $[1\bar{1}]$ direction, as indicated by the red arrow.}\label{fig:illu}
\end{figure}

In recent years, a new class of magnetic materials has emerged, known as altermagnets~\cite{smejkal_beyond_2022}. These materials exhibit collinear-compensated magnetic order, which goes beyond the traditional binary classification of ferromagnets and antiferromagnets. In altermagnets, the opposite spin sublattices are connected through rotation rather than inversion or translation, leading to non-relativistic anisotropic spin splitting in the Brillouin zone. Experimentally, altermagnetism has been found in both metallic materials such as RuO$_2$~\cite{ahn_antiferromagnetism_2019,smejkal_beyond_2022} and Mn$_5$Si$_3$~\cite{reichlova_macroscopic_2021}, and insulating materials such as MnF$_2$~\cite{yuan_giant_2020,smejkal_crystal_2020} and MnTe~\cite{gonzalez_betancourt_spontaneous_2023}. The altermagnets have a unique spin splitting, leading to a wide range of fascinating phenomena~\cite{ma_multifunctional_2021,bai_efficient_2023,das_transport_2023,ghorashi_altermagnetic_2023,ouassou_josephson_2023,papaj_andreev_2023,sun_andreev_2023,zhou_crystal_2023,zhu_topological_2023,zhang_finite-momentum_2023,beenakker_phase-shifted_2023}, such as Andreev reflection~\cite{sun_andreev_2023}, crystal Hall effect~\cite{smejkal_crystal_2020,zhou_crystal_2023}, finite-momentum Cooper pairs in altermagnet/superconductor heterojunctions~\cite{zhang_finite-momentum_2023}, and topological superconductivity~\cite{zhu_topological_2023,li_majorana_2023,ahn_antiferromagnetism_2019}. Given the features of altermagnets, an interesting question arises: can we utilize the novel altermagnets to create and manipulate the higher-order topological states?

In this work, we make a positive response to this question. Specifically, we design a heterostructure made of a TI and an altermagnet to create and manipulate the higher-order topological states, as illustrated in Fig.~\ref{fig:illu}. By the effective model and edge theory, we find that when the in-plane N\'{e}el vector is around the $[1\bar{1}]$ direction, the original helical edge states protected by TRS are gapped with in-gap states localized at the two corners along the $[11]$ direction, i.e., TCSs, as shown in Fig.~\ref{fig:illu}. Furthermore, by changing the orientation of the N\'{e}el vector, we can effectively manipulate these TCSs. The two TCSs can be moved to the other two corners with the N\'{e}el vector rotated around the $[11]$ direction. Based on first-principles calculations, we propose an experimental setup that involves placing 2D buckled bismuth on the surface of the altermagnetic material MnF$_2$~\cite{yuan_giant_2020,smejkal_crystal_2020} to realize and tune such TCSs. The magnetic proximity effect~\cite{vobornik_magnetic_2011} plays a crucial role in inducing altermagnetism and spin splitting within the 2D TI. We confirm the existence of corner states in the MnF$_2$/Bi/MnF$_2$ sandwich structure and demonstrate the tunability of these TCSs. This intriguing setup provides a new platform for realizing non-Abelian statistics by using TCSs~\cite{kitaev_fault-tolerant_2003,Loss2013,wu_double-frequency_2020,wu_non-abelian_2020}.

\textit{Model.---} We first introduce a first-order TI model defined on a square lattice with the Hamiltonian expressed in momentum space as
\eq{
    H_0(\bfk)=M(\bfk)\sigma_z+A_x\sin k_xs_y\sigma_x-A_y\sin k_ys_x\sigma_x,\label{eq:ti}
}
where $M(\bfk)=(m_0-t_x\cos k_x-t_y\cos k_y)$, and $\sigma_i$ and $s_j$ are Pauli matrices acting on the orbital $(a,b)$ and spin $(\ua,\da)$ degree of freedom,  respectively. The 2D TI protected by TRS $\mathcal{T}=is_y\mathcal{K}$, where $\mathcal{K}$ is the complex conjugate, and has inversion symmetry $\mathcal{P}=\sigma_z$. The $\mathcal{Z}_2$ topological invariants can be obtained from the parity eigenvalue on the time-reversal invariant momentum points  $\Gamma_i$~\cite{fu_topological_2007}.
When $m_0^2-(t_x+t_y)^2<0$ is satisfied, the TI with $\mathcal{Z}_2=1$ has TRS-protected helical edge states, as shown by the blue dashed line in Fig.~\ref{fig:corner} (a).

\begin{figure}[h]
    \centering
    \includegraphics[scale=0.19]{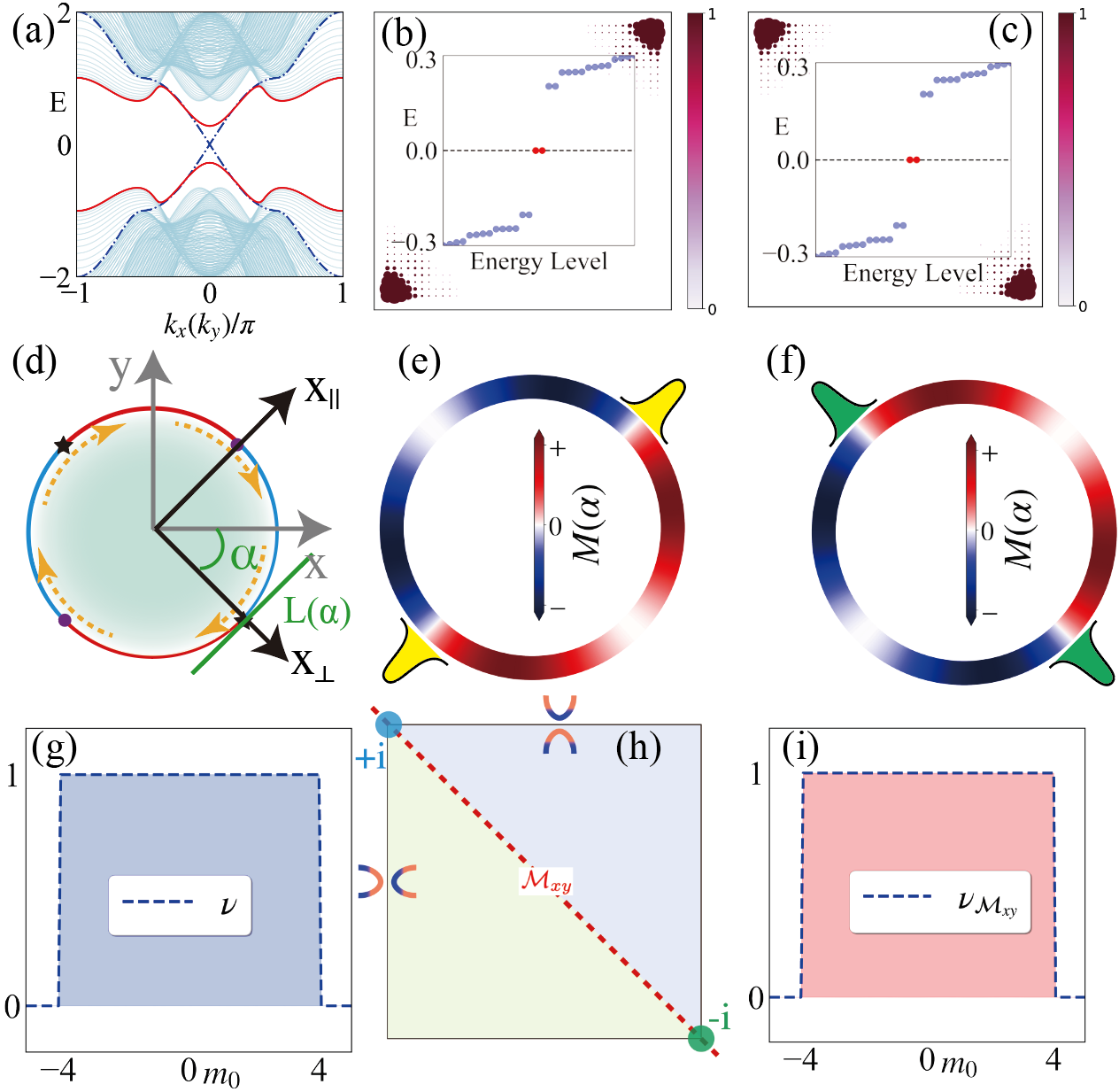}
    \caption{(a) The edge spectrum for a cylinder geometry. The blue dotted lines represent the gapless helical edge states of the topological insulator. The red solid lines denote the gapped helical edge states after the altermagnet is turned on. (b) Inset: Two in-gap states emerge with the N\'{e}el vector along the $[1\bar{1}]$ direction. The real spatial distribution of their wave function is plotted. (c) Same as (b) but with the N\'{e}el vector along the $[11]$ direction. (d) The tangents $L(\alpha)$, on which we will develop the generic edge theory, mark different boundaries with the clockwise rotation angle $\alpha$. (e) (f) The change of the boundary Dirac mass with the rotation angle $\alpha$ when the N\'{e}el vector is along the $[1\bar{1}]$ direction and the $[11]$ direction, respectively.  (g) The topological invariant $\nu$ is plotted as a function of the $m_0$. (h) Schematic diagram: The boundary of the mirror symmetry operation connection has opposite Dirac mass, and the TCSs originate from different mirror subspaces. (i) The mirror-graded winding number $\nu_{\mathcal{M}_{xy}}$ is plotted as a function of the $m_0$.  Common parameters: $m_0=1.0, t_x=t_y=A_x=A_y=2.0,J_0=0.5$.}\label{fig:corner}
\end{figure}

The spin splitting of altermagnets exhibits various forms, including $d$-wave, $g$-wave, and $i$-wave~\cite{smejkal_beyond_2022}. We take the proximity-induced $d$-wave spin splitting as an example which reads
\eq{
    H_{\rm AM}(\bfk)=2J_0(\cos k_x-\cos k_y)\bsf{s}\cdot\bsf{\hat{n}},
}
where the vector $\bsf{\hat{n}}=(\sin\theta\cos\varphi,\sin\theta\sin\varphi,\cos\theta)$ represents the direction of the N\'{e}el vector. Here the in-plane N\'{e}el component ($\theta=\pi/2$) is considered with the out-of-plane component left in the Supplemental Material (SM)~\cite{supp2}. As shown in Fig.~\ref{fig:corner}(a), the original gapless helical edge states develop a gap when the in-plane N\'{e}el vector aligns along the $[1\bar{1}]$ direction. We further calculate the energy spectrum of a finite-size square sample, as displayed in the inset of Fig.~\ref{fig:corner}(b). One can observe that two in-gap states emerge in the edge gap. We plot the wave function distribution of these in-gap states and find them to be localized at two corners of the square sample, as depicted in Fig.~\ref{fig:corner}(b), which means the two localized in-gap states are possible TCSs. This provides evidence for the presence of the HOTI state in the system when altermagnetism is activated, indicating a topological phase transition from a first-order TI to a HOTI state.

When the N\'{e}el vector is directed along the $[11]$ direction, the system also exhibits the HOTI states with the hallmark TCSs but on the other corners, as depicted in Fig.~\ref{fig:corner} (c).  Notice that more TCSs can be obtained with N\'{e}el vector along other directions or by altermagnets with higher angular momentum quantum numbers, such as $g$-wave and $i$-wave~\cite{supp2}. Therefore, by rotating the orientation of the N\'{e}el vector, we offer an effective method to manipulate the TCSs, enabling dynamic control and repositioning of these states within the system. In practical experiments, the orientation of the N\'{e}el vector can be controlled by applying an electric field or a spin-orbit torque~\cite{meinert_electrical_2018-1,godinho_electrically_2018,bodnar_writing_2018}. Thus, this proposal opens up new possibilities for realizing non-Abelian statistics of Dirac fermions with fractional charge~\cite{wu_non-abelian_2020}.

\textit{Symmetry analysis and edge theory.---} Our proposed model exhibits $C_{2z}T=s_x\mathcal{K}$ symmetry and falls within the Stiefel-Whitney (SW) class with its topology characterized by the second SW number $w_2$~\cite{bouhon_wilson_2019,ahn_symmetry_2019}. Applying a unitary transformation $U=\exp(i\pi/4s_x)$ to $H(\bfk)=H_0(\bfk)+H_{\rm AM}(\bfk)$, a real form can be obtained~\cite{supp2}. We calculate the SW number $w_2$ by using the Wilson loop method with $w_2=1$, signifying a non-trivial HOTI~\cite{supp2}.

The centrosymmetric topological insulator we consider has particle-hole symmetric energy bands and thus can be considered to have chiral symmetry $\mathcal{C}$ approximately~\cite{ren_engineering_2020,chen_universal_2020-1}.  The introduction of altermagnetsim into the topological insulator breaks $\mathcal{T}$, while $\mathcal{P}$ and $\mathcal{C}$ remain intact. Such higher-order topological states belong to the $\mathbb{Z}_2$ classification~\cite{khalaf_higher-order_2018}. The topological invariant characterizing the higher-order topology reads as
$\nu=\sum_{\Gamma_i}n^{-}(\Gamma_i)/2\quad \text{mod}\quad 2$,
where $n^{-}(\Gamma_i)$ is the number of occupied states with negative parity eigenvalue at time-reversal invariant points $\Gamma_i$.
We calculate the topological invariant $\nu$, as shown in Fig.~\ref{fig:corner}(g), and $\nu=1$ with $m_0\in\cs{-4,4}$ indicates a HOTI.

The origin of these TCSs and their tunability in real space as the N\'{e}el vector changes can be understood through the edge theory. We use the Hamiltonian $H(\bfk)=H_0(\bfk)+H_{\rm AM}(\bfk)$ to describe the HOTI. Expanding $H(\bfk)$ at $\bsf{\Gamma}=(0,0)$ to the second order yields
\eqa{
    H^{\rm eff}(\bfk)&=(m+\frac{t_x}{2}k_x^2+\frac{t_y}{2}k_y^2)\sigma_z+A_xk_x\sigma_xs_y\\
    &-A_yk_y\sigma_xs_x-J_0(k_x^2-k_y^2)\bsf{s}\cdot\bsf{\hat{n}},
}
where $m=m_0-t_x-t_y$.  We consider an arbitrary boundary $L(\alpha)$, which is the tangent with the clockwise rotation angle $\alpha$, as shown in Fig.~\ref{fig:corner}(d). The coordinate axes need to be rotated to obtain the new momentum $k_\parallel$ and $k_\perp$. In the new coordinates, when the strength of the altermagnets is smaller than the bulk gap, the Hamiltonian can be decomposed as $H^{\rm eff}(\bfk)=H_0(\bfk)+H_p(\bfk)$ (see details in the SM~\cite{supp2}). Consider the semi-infinite plane $x_\perp\in(-\infty,0]$, where a boundary exists at $x_\perp=0$.  The momentum $k_\perp$ is replaced by $-i\partial_\perp$ and the eigenequation $H_0\psi_\alpha(x_\perp)=E_\alpha\psi(x_\perp)$ is solved with the boundary condition $\psi(0)=\psi(-\infty)=0$. Two solutions for $E_\alpha=0$ are obtained with $\psi_\alpha(x_\perp)=\mathcal{N}_\perp\sin(\kappa_1x_\perp)e^{\kappa_2x_\perp}e^{ik_\parallel x_\parallel}\chi_\alpha$, where the normalization constant is given by $|\mathcal{N}_\perp|^2=4|\kappa_2(\kappa_1^2+\kappa_2^2)/\kappa_1^2|$ and the eigenvector $\chi_\alpha$ satisfies $(\sin\alpha s_y+\cos\alpha s_x)\sigma_y\xi=\xi$. We  choose $\chi_i$ as $\chi_1=1/\sqrt{2}(-i e^{-i \alpha }, 0 , 0 ,1)^T$ and $\chi_2=1/\sqrt{2}(0,i e^{-i \alpha } ,1 , 0)^T$ and project perturbation $H_p$ onto the  bases $\cs{\psi_1,\psi_2}$, and obtain the boundary Hamiltonian for any boundary $L(\alpha)$ and any N\'{e}el vector with polar $\theta$ and azimuthal $\varphi$ angles
\eq{
    H_{\rm eff}(x_\perp,k_\parallel)=Ak_\parallel\eta_z+M(\alpha,\theta,\varphi)\eta_y,
}
where $\eta_i$ are Pauli matrices acting on $\psi_i$.

The Dirac mass term that arises from altermagnets is given by
\eq{
    M(\alpha,\theta,\varphi)\sim J_0\sin\theta\cos(2\alpha)\cos\cs{\varphi-\alpha}.
}
Our research primarily concentrates on the in-plane component of the N\'{e}el vector with $\theta = \pi/2$, while the out-of-plane component of the N\'{e}el vector does not influence the edge states (see details in the SM~\cite{supp2}).
\begin{figure*}[t]
    \centering
    \includegraphics[scale=0.25]{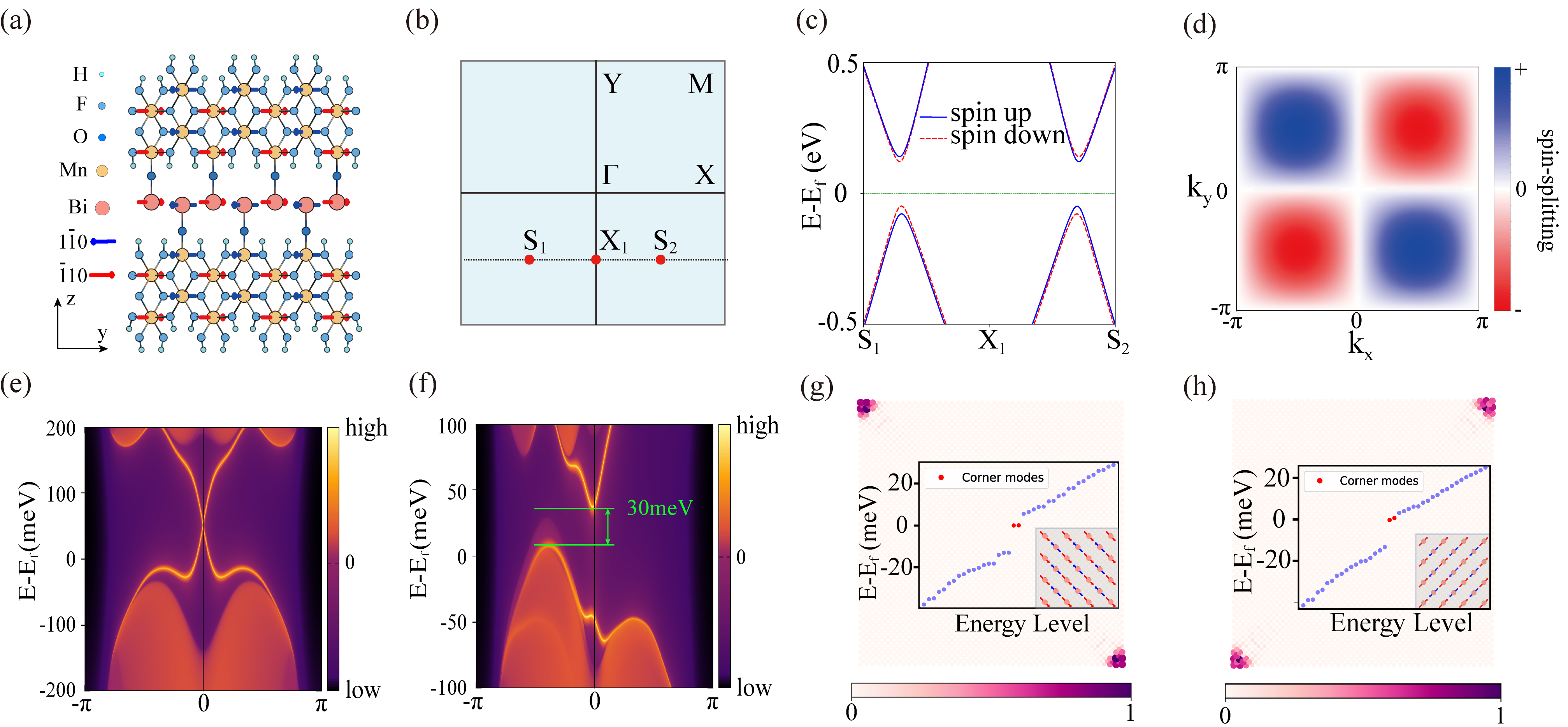}
    \caption{(a) Side view of the optimized crystal structure of the MnF$_{2}$/Bi/MnF$_{2}$ configuration. (b) The 2D Brillouin zone with high-symmetric points and lines. (c) The spin-polarized energy bands of the MnF$_{2}$/Bi/MnF$_{2}$ sandwich structure, where the blue-solid/red-dashed lines represent the spin-up/spin-down. (d) The $d$-wave spin-polarized band splitting induced in bismuthene from our DFT calculation. (e) Helical edge states are calculated via DFT in the absence of altermangetism. (f) Gapped edge states obtained by the DFT calculation with the N\'{e}el vector aligned along the $[1\bar{1}]$ direction. (g) and (h) N\'{e}el vector along the $[1\bar{1}]$ and $[11]$ directions, respectively. Inset: Two in-gap topological corner states emerge. The small arrows depict the direction of the N\'{e}el vector. The real spatial distribution of their wave function is plotted.}\label{fig:material}
\end{figure*}
For the scenario that the N\'{e}el vector is along the $[\bar{1}1]$ direction ($\varphi=3\pi/4$), we calculate and plot the Dirac mass at any edge $L(\alpha)$, as depicted in Fig.~\ref{fig:corner} (e). One can observe that at the clockwise rotation angles of $\alpha=3\pi/4$ and $7\pi/4$, there exist domains in the Dirac mass that host zero-energy bound states resembling Jackiw-Rebbi zero modes~\cite{jackiw_solitons_1976}, which is consistent with the numerical results of a square sample shown in Fig.~\ref{fig:corner} (b). Although the Dirac mass $M(\alpha)$ vanishes at $\alpha=\pi/4$ and $5\pi/4$, the lack of a mass domain prevents the formation of the TCSs. When the N\'{e}el vector is aligned with the $[11]$ direction, two mass domains are formed at $\alpha=\pi/4$ and $5\pi/4$, respectively, as plotted in Fig.~\ref{fig:corner} (f). The two TCSs can be moved by rotating the N\'{e}el vector. Consequently, we can not only create TCSs but also manipulate them by an altermagnet.

It is worth noting that when the N\'{e}el vector is along the $[11]$ direction, the system has extra mirror symmetry $\mathcal{M}_{xy}=i\sqrt{2}/2(s_x+s_y )$ along a line defined by $k_x=-k_y$. Along the mirror-invariant line, the Hamiltonian $H(\bfk)=H_0(\bfk)+H_{\rm AM}(\bfk)$ can be decomposed into distinct mirror subspaces labeled by the  $\pm i$. In each subspace, the Hamiltonian expressed as $\mathcal{H}^{\pm i}=\bsf{q}_{\pm i}(k)\cdot\bsf{\sigma}$, which are two 1D SSH models with opposite winding number $\nu_{+i}=-\nu_{-i}$~\cite{supp2}. Intuitively, the non-trivial SSH model will result in two end states at an endpoint. However, when the N\'{e}el vector deviates from $[11]$, mirror symmetry $\mathcal{M}_{xy}$ is broken,  and the gap of the system is maintained. This indicates that regardless of the presence or absence of $\mathcal{M}_{xy}$, the system is in the same topological phase and has the $\mathbb{Z}_2$ classification~\cite{khalaf_higher-order_2018,langbehn_reflection-symmetric_2017}, only one end state stable at one corner.  As a result, the system will only exhibit two corner states instead of four. Furthermore, the chiral symmetry $\mathcal{C}=s_z\sigma_x$ which satisfies $\cl{\mathcal{M}_{xy}, \mathcal{C}}=0$ implies that the two corner zero modes originate from different mirror subspaces, as shown in Fig.~\ref{fig:corner}(h). In the presence of mirror symmetry, the non-trivial second-order topology can currently be characterized by a mirror-graded winding number, which is defined as $\nu_{\mathcal{M}_{xy}}=(\nu_{+i}-\nu_{-i})/2$. A nonzero $\nu_{\mathcal{M}_{xy}}$ indicates that the system has a nontrivial second-order topology~\cite{bercioux_topological_2018}. The calculated $\nu_{\mathcal{M}_{xy}}$  is shown in Fig.~\ref{fig:corner}(i), which is consistent with  Fig.~\ref{fig:corner}(g), confirming the non-trivial topology of the system. The investigation of the N\'{e}el vector along $[1\bar{1}]$ is similar~\cite{supp2}.

The TCSs still robustly exist when the N\'{e}el vector deviates from the $[11]$ or $[1\bar{1}]$ direction. As long as the condition $M(\alpha,\pi/2,\varphi) M(\alpha+\pi/2,\pi/2,\varphi)<0$ or $M(\alpha,\pi/2,\varphi) M(\alpha-\pi/2,\pi/2,\varphi)<0$ is met, the system will have TCSs. Since $M(\alpha+\pi/2,\pi/2,\varphi)=-M(\alpha-\pi/2,\pi/2,\varphi)$, the condition is equivalent to $M(\alpha,\pi/2,\varphi) M(\alpha+\pi/2,\pi/2,\varphi)\neq0$, i.e., $\varphi \notin \left\{3\pi/2-\alpha,\pi-\alpha,\pi/2-\alpha,\alpha \right\} $. In principle, TCSs will exist as long as the condition holds.  However, along these two directions $\varphi=\pi/4-\alpha$ and $\varphi=3\pi/4-\alpha$, corresponding to $[11]$ or $[1\bar{1}]$ directions with $\alpha=0$, the edge gap is large, which is convenient for experimental observation of TCSs.

We also derive the general boundary Hamiltonian for the other anisotropic spin splittings with non-zero angular momentum quantum numbers, such as $g$-wave and $i$-wave, and the isotropic spin splitting of $s$-wave with zero angular momentum. The previous proposals to induce TCSs by using the Zeeman field can be considered as the special isotropic $s$-wave case in our proposal~\cite{supp2}.

\textit{Material realization.---} Based on first-principles calculations, as a demonstration, we propose that a MnF$_2$/Bi/MnF$_2$ sandwich with optimized structure depicted in Fig.~\ref{fig:material} (a) can achieve such tunable TCSs (see details in the SM~\cite{supp2}). Previous studies have shown the potential of two-dimensional bismuthene, with its honeycomb lattice structure, to be used as a material for TIs~\cite{songQuantumSpinHall2014,zhou_epitaxial_2014,liu_low-energy_2014,hsuNontrivialElectronicStructure2015,reis_bismuthene_2017}. And bismuthene has been successfully synthesized on SiC substrates~\cite{reis_bismuthene_2017}. To ensure lattice alignment between bismuthene and MnF$_2$, we utilize bismuthene with a buckled square lattice structure. This specific configuration of bismuthene, as a first-order TI, has an energy gap of approximately $0.69$ eV~\cite{luoRoomTemperatureQuantum2015}, making it suitable for observing the TCSs. Through the magnetic proximity effect~\cite{vobornik_magnetic_2011}, magnetism can be induced in the buckled bismuthene.   We employ a sandwich structure with $[C_2][S_4]$ symmetry where $S_4$ connects the sublattices with opposite spin in real space and $C_2$ inverse the spins in spin space~\cite{smejkal_beyond_2022}, ensuring that the magnetic moment produced through the proximity effect satisfies the restriction of altermagnetism, as illustrated in Fig.~\ref{fig:material} (a).   Based on first-principles density functional theory (DFT) calculations, we present the spin-polarized energy bands in Fig.~\ref{fig:material}(c) (see details in the SM~\cite{supp2}). Notably, along a horizontal line ($S_1X_1S_2$) in the Brillouin zone, as shown in Fig.~\ref{fig:material} (b), the energy band exhibits the unique spin splitting with spin-up/down in blue-solid/red-dashed lines, demonstrating the characteristic behavior of $d$-wave spin splitting, as shown in Fig.~\ref{fig:material} (d). This spin splitting, approximately $30$ meV, confirms the effective induction of altermagnetism in bismuthene through the magnetic proximity effect.

To better investigate the electronic structure and topological properties, we construct an \textit{ab initio} tight-binding model based on DFT and Wannier function~\cite{marzari_maximally_1997-1,souza_maximally_2001}.  The model can take into account the effect of altermagnetism in MnF$_2$.   Additionally, we build a minimal tight-binding model based on symmetry to capture the physics of the DFT results~\cite{supp2}.   In the absence of altermagnetism, we observe gapless edge states within the bulk gap of the buckled bismuthene, as shown in Fig.~\ref{fig:material} (e). However, by activating the altermagnetism of MnF$_2$ with the in-plane N\'{e}el vector  along the diagonal direction, the helical edge states acquire a gap of approximately $30$ meV, as shown in Fig.~\ref{fig:material} (f). A similar phenomenon is observed when the N\'{e}el vector  along the off-diagonal direction~\cite{supp2}. These indicate the breakdown of the first-order topology of the system and possibly the higher-order topology induced by the altermagnet. We calculate the energy spectrum of a finite-size square sample of the MnF$_2$/Bi/MnF$_2$ sandwich structure, as shown in Figs.~\ref{fig:material} (g) and (h). When the N\'{e}el vector has an in-plane component, it can be observed that two in-gap states emerge in red and the corresponding distribution of wave function is localized at two corners of the sample. Consequently, despite the altermagnetism causing the first-order topology of the system to be trivial, the system exhibits a non-trivial second-order topology. Moreover, the position of the corner states can be manipulated by rotating the N\'{e}el vector. This observation is consistent with our theoretical model results.  Additionally, upon enabling spin-orbit coupling, the sandwich structure exhibits $C_{2z}T$ symmetry with its nontrivial second SW number $w_2$ confirmed via the Wilson loop method~\cite{supp2}. Therefore, by employing DFT calculations, we demonstrate the capability to achieve and modulate the TCSs through manipulation of the N\'{e}el vector direction and provide a candidate material setup.

\textit{Discussion.---}Motivated by the unique spin-polarized band splitting in altermagnetic materials, we propose a route to create and manipulate TCSs by altermagnets. Our DFT calculations have confirmed that the magnetic proximity effect induces a spin-polarized band splitting in TIs and suggest a MnF$_2$/Bi/MnF$_2$ sandwich structure as a candidate material setup for realizing our proposal.  The braiding of TCSs can be facilitated by constructing a cross-shaped or $T$-shaped geometry~\cite{alicea_non-abelian_2011}. A gate electrode is positioned at the intersection point, allowing the generation and fusion of TCSs to be achieved by controlling the channel's opening and closing through gate voltage control~\cite{wu_double-frequency_2020}. Our proposal expands the possibility of TCS manipulation by controlling the N\'{e}el vector, which can employ gate voltage control to enable more intricate braiding operations and pave the way for topological quantum computing.   In the experiment, the direction of the N\'{e}el vector in altermagnets like other antiferromagnetic materials can be manipulated and detected using techniques such as current, voltage~\cite{godinho_electrically_2018,mahmood_voltage_2021}, strain~\cite{parkStrainControlNeel2019, shiraziStressinducedNeelVector2022}, and spin-orbit torques~\cite{zhang_control_2022}. Moreover, altermagnetic materials exhibit robustness to external magnetic field perturbations, and their ultrafast response due to the absence of a coercivity field~\cite{baltz_antiferromagnetic_2018}. These fascinating properties open up the exciting possibility of achieving non-Abelian braiding of Dirac fermions~\cite{Loss2013,wu_non-abelian_2020,wu_double-frequency_2020}, providing new avenues for future research and technological applications.

The existence of charged TCSs can be detected using scanning tunneling microscopy (STM)~\cite{yin_probing_2021}. Additional evidence for the presence of TCSs can be provided by exploiting the ability of the N\'{e}el vector to modulate the topological states. In the case where the N\'{e}el vectors are oriented along the $z$ axis ($\theta=0$), the system has gapless edge states. When the energy of the STM probe approaches that of the edge states, a broadening peak can be observed along the boundary. By adjusting the N\'{e}el vector to open a gap in the edge state, sharp peaks can be observed at the corners of the sample using STM. Consequently, the observation of changing energy spectrum peaks with STM can serve as strong evidence for the existence of corner states. Furthermore, by manipulating the N\'{e}el vector, it is possible to control the movement of the corner states along the boundary and achieve a quantum charge pump independently of the bulk state~\cite{wu_quantized_2022}. This effect can be precisely measured through transport experiments, further supporting the understanding of the influence of the N\'{e}el vector on the topological properties of the system. The ability to control the movement of corner states and achieve a quantum charge pump holds potential implications for applications in quantum information processing and topological devices.

\textit{Note added.—}Recently, there appeared a related preprint~\cite{ezawa_detecting_2024} that focuses on the altermagnets and higher-order topology.

    \textit{Acknowledgments.}---It is our pleasure to thank Shifeng Qian and Yongpan Li for their insightful discussions. The work is supported by the National Key $\&$ Program of China (Grant No. 2020YFA0308800), the NSF of China (Grant No. 12374055), and the Science Fund for Creative Research Groups of NSFC (Grant No. 12321004).  Yu-Xuan Li and Yichen Liu contribute equally to the work.
\bibliography{ref}

\clearpage
\onecolumngrid
\begin{center}
    \textbf{\large Supplementary material for "Creation and Manipulation of Higher-Order Topological States by Altermagnets"}\\[.2cm]
    Yu-Xuan Li, Yichen Liu and  Cheng-Cheng Liu \\[.1cm]
    {\itshape Centre for Quantum Physics, Key Laboratory of Advanced Optoelectronic Quantum Architecture and Measurement (MOE), School of Physics, Beijing Institute of Technology, Beijing 100081, China}
\end{center}

\maketitle
\setcounter{equation}{0}
\setcounter{figure}{0}
\setcounter{table}{0}
\setcounter{page}{1}
\renewcommand{\theequation}{S\arabic{equation}}

\renewcommand{\thefigure}{S\arabic{figure}}
\renewcommand{\thetable}{\arabic{table}}
\renewcommand{\tablename}{Supplementary Table}

\renewcommand{\bibnumfmt}[1]{[S#1]}
\renewcommand{\citenumfont}[1]{#1}
\makeatletter

\maketitle

\setcounter{equation}{0}
\setcounter{section}{0}
\setcounter{figure}{0}
\setcounter{table}{0}
\setcounter{page}{1}
\renewcommand{\theequation}{S-\arabic{equation}}

\renewcommand{\thefigure}{S\arabic{figure}}
\renewcommand{\thetable}{\arabic{table}}
\renewcommand{\tablename}{Supplementary Table}

\renewcommand{\bibnumfmt}[1]{[S#1]}
\makeatletter

\maketitle

This supplementary material is divided into six sections. Section \ref{sec:edge} provides the edge theory derivation with the N\'{e}el vector along an arbitrary direction and demonstrates the manipulation of the topological corner states by rotating the N\'{e}el vector. Section~\ref{sec:mgn} includes mirror-graded winding number  when N\'{e}el vector
along $[1\bar{1}]$ direction. Section~\ref{edge-gi} provides the general edge Hamiltonian for $g$-wave, $i$-wave as well as $s$-wave spin splitting. Section \ref{sec:dft} contains the computational details of the DFT. Section \ref{sec:sand} introduces a sandwich structure consisting of a 2D TI and altermagnets. In Section \ref{sec:tb}, an effective model is constructed using DFT and Wannier functions, and the manipulation of the corner states by changing the direction of the N\'{e}el vector is verified. In Section~\ref{sec:tb_model}, a minimal tight-binding model was constructed based on the symmetry of the sandwich structure, which well simulates the modulation of the N\'{e}el vector on the corner states. In Section~\ref{sec:sw}, the topological classification of the system is discussed and the second Stiefel-Whitney number $w_2$ is determined using the Wilson loop method.

\section{Edge Theory}\label{sec:edge}
\begin{figure}[h]
    \centering
    \includegraphics[scale=0.5]{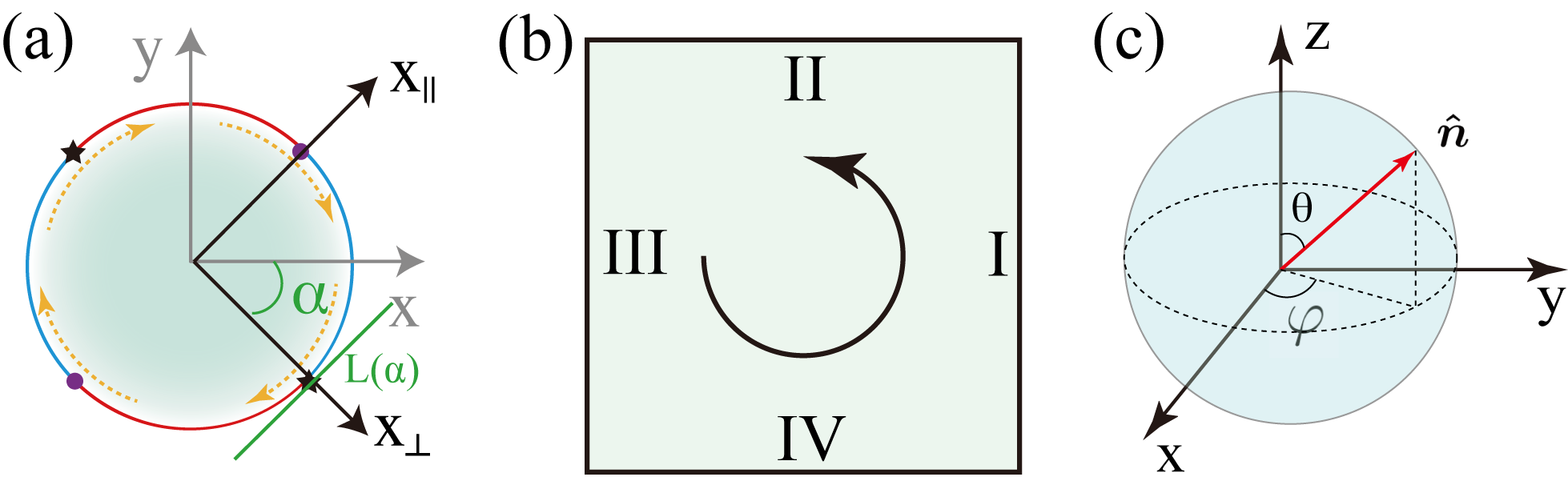}
    \caption{(a) Schematic diagram of rotation of the coordinate system. The tangent line L($\alpha$) represents any boundary with the rotated coordinates ($x_{||}$,$x_\bot$) and an angle $\alpha$ to the horizontal axis. (b) The schematic diagram illustrates the edge coordinate $\ell$ and defines the counterclockwise direction as the positive direction. The four edges, labeled I, II, III, and IV, are delineated for application in the edge theory. (c) The unit vector $\bsf{\hat{n}}$ of the N\'{e}el vector with the polar angle $\theta$ and azimuthal $\varphi$ angles in the spherical coordinates. }\label{fig:circle}
\end{figure}

In this section, we investigate the edge theory for the $d$-wave case with the N\'{e}el vector oriented in any direction. Figure~\ref{fig:circle}(c) visually depicts this scenario, where $\hat{\bsf{n}}$ denotes the N\'{e}el vector, $\theta$ represents the polar angle, and $\varphi$ represents the azimuthal angle.

Firstly, our focus is on the scenario wherein solely in-plane N\'{e}el vector components are present, i.e. azimuth angle $\theta=\pi/2$. The low-energy Hamiltonian at the $\bsf{\Gamma}$ point reads as follows
\eq{
    H(\bfk)=\cs{m+\frac{t_x}{2}k_x^2+\frac{t_x}{2}k^2_y}\sigma_z+A_0(k_xs_y-k_ys_x)\sigma_x-J_0(k_x^2-k_y^2)(\sin\theta\cos\varphi s_x+\sin\theta\sin\varphi s_y),\label{eq:eff1}
}
where $m=m_0-t_x-t_y, J_0^x=J_0\sin\theta\cos\varphi$ and $ J_0^y=J_0\sin\theta\sin\varphi $. We consider an arbitrary edge $L(\alpha)$, which needs to rotate the coordinate axis clockwise to obtain new momentum $k_\parallel$ and $k_\perp$, as shown in Fig.~\ref{fig:circle}(a). The relationship between the coordinates before and after the rotation can be described as follows
\eq{
    \cs{
        \begin{array}{c}
            k_x\\
            k_y
        \end{array}
    }=\cs{
        \begin{array}{cc}
            \cos\alpha&-\sin\alpha\\
            \sin\alpha&\cos\alpha
        \end{array}
    }\cs{
        \begin{array}{c}
            k_\parallel\\
            k_\perp
        \end{array}
    }.
}

In the new coordinate system, the low-energy Hamiltonian, as given by Eq.~\eqref{eq:eff1}, can be expressed as follows
\eqa{
    H(k_\parallel,k_\perp)&=M(k_\parallel,k_\perp)\sigma_z+A\cm{k_\parallel\cs{\cos\alpha s_y-\sin\alpha s_x}-k_\perp\cs{\sin\alpha s_y+\cos\alpha s_x}}\sigma_x\\
    &-J^x_0\cm{\cos2\alpha(k_\perp^2-k_\parallel^2)+2\sin2\alpha k_\parallel k_\perp}s_x\\
    &-J^y_0\cm{\cos2\alpha(k_\perp^2-k_\parallel^2)+2\sin2\alpha k_\parallel k_\perp}s_y.
}
When the strength of the altermangetism is small  compared to the bulk gap, the Hamiltonian can be divided into two parts, $H_0$ and $H_p$
\eqa{
    H_0(k_\parallel,k_\perp)&=M(k_\parallel,k_\perp)\sigma_z-Ak_\perp\cs{\sin\alpha s_y+\cos\alpha s_x}\sigma_x,\\
    H_p(k_\parallel,k_\perp)&=-J^x_0\cm{\cos2\alpha(k_\perp^2-k_\parallel^2)+2\sin2\alpha k_\parallel k_\perp}s_x-J^y_0\cm{\cos2\alpha(k_\perp^2-k_\parallel^2)+2\sin2\alpha k_\parallel k_\perp}s_y\\
    &+Ak_\parallel\cs{\cos\alpha s_y-\sin\alpha s_x}.
}
Consider the semi-infinite system bounded by $x_\perp=0$ and occupying the space $x_\perp\in(-\infty,0]$. We begin by solving the eigenequation $H_0(k_\parallel=0,-i\partial_\perp)\psi(x_\perp)=E\psi(x_\perp)$. When the boundary condition $\psi(0)=\psi(-\infty)=0$ is satisfied, one can obtain two zero-energy solutions
\eq{
    \psi_\alpha(x_\perp)=\mathcal{N}_\perp\sin(\kappa_1x_\perp)e^{\kappa_2x_\perp}e^{ik_\parallel x_\parallel}\xi_\alpha,
}
where the normalization parameter is given by $\cabs{\mathcal{N}_\perp}^2=4\cabs{\kappa_2(\kappa_1^2+\kappa_2^2)/\kappa_1^2}$. The parameters $\kappa_1$ and $\kappa_2$ can be expressed as follows
\eq{
    \kappa_1=\sqrt{\cabs{\frac{2m}{t_0}}-\frac{A^2}{t_0^2}},\qquad \kappa_2=\frac{A}{t_0}.
}
The $\xi_\alpha$ satisfy eigenequation $(\sin\alpha s_y+\cos\alpha s_x)\sigma_y\xi_\alpha=\xi_\alpha$. One can obtain the following solutions
\eq{
    \xi_1=\frac{1}{\sqrt{2}}\left(
    \begin{array}{c}
        -i e^{-i \alpha } \\
        0 \\
        0 \\
        1 \\
    \end{array}
    \right),\qquad \xi_2=\frac{1}{\sqrt{2}}\left(
    \begin{array}{c}
        0 \\
        i e^{-i \alpha } \\
        1 \\
        0 \\
    \end{array}
    \right).
}
\begin{figure}[t]
    \centering
    \includegraphics[scale=0.45]{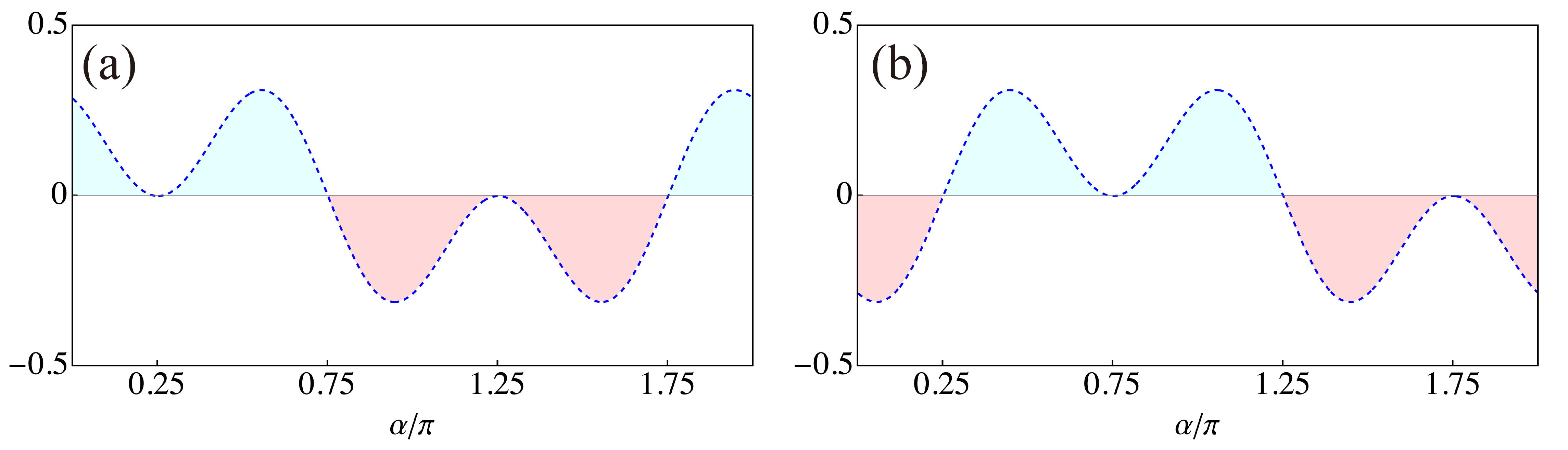}
    \caption{The variation of the Dirac mass term for the boundary direction $\alpha$ induced by an in-plane altermagnets. (a) The N\'{e}el vector is oriented along the $[1\bar{1}]$ direction, while in (b), it is along the $[11]$ direction.}\label{fig:mass}
\end{figure}
Subsequently, the $H_p$ is considered as perturbations and projected onto the basis vectors $\xi_{1,2}$ to yield
\eqa{
    &\kb{\xi_i|Ak_\parallel(\cos\theta s_y-\sin\theta s_x)\sigma_x|\xi_j}=Ak_\parallel\left(
    \begin{array}{cc}
        1& 0 \\
        0 & -1 \\
    \end{array}
    \right),\\
    &\kb{\xi_i|-J^x_0(\cos(2\alpha)(k_\perp^2-k_\parallel^2))s_x\sigma_0|\xi_j}=-J^x_0\cos(2\alpha)\cos(\alpha)\left(
    \begin{array}{cc}
        0 & i \\
        -i & 0 \\
    \end{array}
    \right)k_\perp^2,\\
    &\kb{\xi_i|-J^y_0(\cos(2\alpha)(k_\perp^2-k_\parallel^2))s_y\sigma_0|\xi_j}=-J^y_0\cos(2\alpha)\sin(\alpha)\left(
    \begin{array}{cc}
        0 & i \\
        -i & 0 \\
    \end{array}
    \right)k_\perp^2.\label{eq:ed2}
}
From Eq.~\eqref{eq:ed2}, the edge  Hamiltonian  can be expressed as
\eqa{
    H_{\rm eff}(x_\perp,k_\parallel)=Ak_\parallel\eta_z+M(\alpha,\theta,\varphi)\eta_y,
}
where the Dirac mass term reads as
\eqa{
    M(\alpha,\theta,\varphi)&\sim\cos(2\alpha)\cm{J_0\sin\theta\cos\varphi\cos(\alpha)+J_0\sin\theta\sin\varphi \sin(\alpha)}\label{eq:diracmass}.
}
When the magnetization in the in-plane N\'{e}el order along the $[11]$ direction, characterized by $\theta=\pi/2$ and $\varphi=\pi/4$, the Dirac mass term contribution depends on the edge direction $\alpha$. This dependence is illustrated in Fig.~\ref{fig:mass}(b), where the Dirac mass terms near $\alpha=3\pi/4$ and $7\pi/4$ exhibit opposite signs. This sign difference gives rise to the emergence of corner states localized at the boundaries between regions with opposite signs of mass. In the case of the magnetization in the in-plane N\'{e}el order along the $[\bar{1}1]$ direction, i.e.,  $\theta=\pi/2$ and $\varphi=\pi/4$, the mass domain wall occurs at $\alpha=\pi/4$ and $5\pi/4$, coinciding with the positions of the corner states, as shown in Fig.~\ref{fig:mass}(a).

\begin{figure}[h]
    \centering
    \includegraphics[scale=0.48]{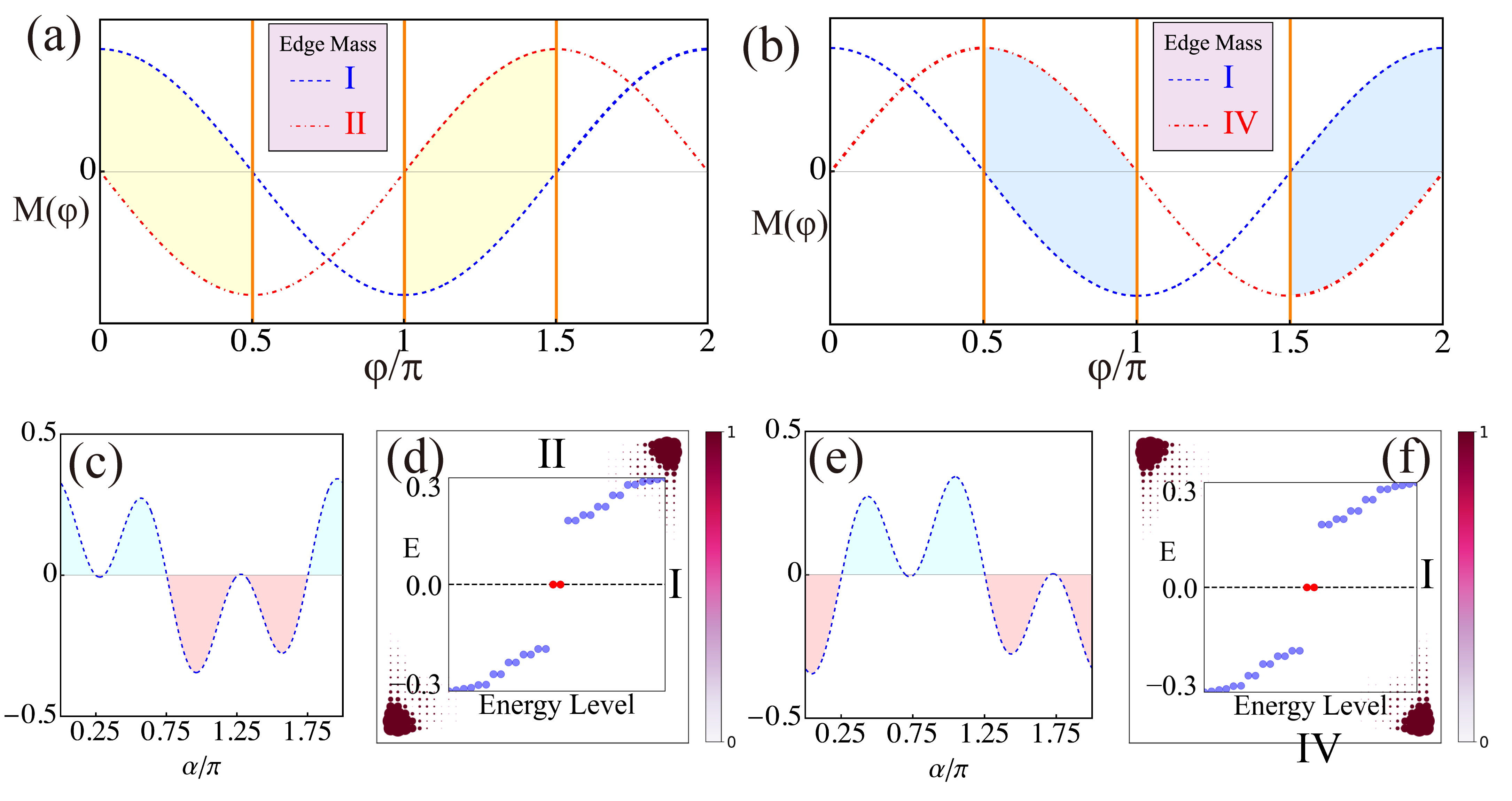}
    \caption{(a) The Dirac masses of boundaries I and II are displayed as a function of the azimuth angle $\varphi$. (b) The Dirac masses of boundaries I and IV are shown. For (c) and (d), the variation of the Dirac mass term induced by in-plane altermagnets is depicted for the boundary direction $\alpha$. (c) The azimuth angle of the N\'{e}el vector is set to $\varphi=0.8\pi$. (e) The azimuth angle is $\varphi=0.2\pi$. The real-space energy spectrum and wavefunction distribution of zero-energy modes are examined for two scenarios: (d) with $\varphi=0.8\pi$, (f) with $\varphi=0.2\pi$.}\label{fig:neelshift}
\end{figure}

It is important to note that even when the  N\'{e}el order deviates from alignment with the $[11]$ or $[\bar{1}1]$ direction, the presence of a domain wall persists. Considering the square geometric structure, it is convenient to designate each boundary segment as I, II, III, and IV, corresponding to the angles $\alpha$ of $0$, $\pi/2$, $\pi$, and $3\pi/2$, respectively, as illustrated in Fig.~\ref{fig:circle}(b). By utilizing Eq.~\eqref{eq:diracmass}, one can calculate the contribution to the Dirac mass term associated with each edge. Importantly, when the Dirac mass contributions on adjacent boundaries  exhibit opposite signs, satisfying
\eq{
    M\cs{\alpha_i,\frac{\pi}{2},\varphi}\times M\cs{\alpha_j,\frac{\pi}{2},\varphi}<0,\quad i=\text{I},\text{III};~j=\text{II},\text{IV},
}
a Jackiw-Rebbi-like bound state emerges at the corner~\cite{jackiw_solitons_1976}, i.e., corner states. Figures~\ref{fig:neelshift}(a) and (b) illustrate the dependence of the Dirac mass contributions along boundaries I, II, and IV on the azimuth angle $\varphi$.
It has been observed that a  mass domain emerges at the intersection of boundaries I and II when the azimuth angle $\varphi$ satisfies the condition
\eq{
    \varphi\in\cs{0,\frac{\pi}{2}}\cup\cs{\pi,\frac{3\pi}{2}}\label{eq:varphi}.
}
Consequently, whenever the azimuth angle of the N\'{e}el vector satisfies Eq.~\eqref{eq:varphi}, a corner state manifests at the corner along the $[1\bar{1}]$ direction. Additionally, when the rotation of the N\'{e}el vector satisfies the condition
\eq{
    \varphi\in\cs{\frac{\pi}{2},\pi}\cup\cs{\frac{3\pi}{2},2\pi},
}
a distinct mass domain emerges at the intersection of boundaries I and IV, as depicted in Fig.~\ref{fig:neelshift}(b).  Taking the azimuth angle $\varphi=0.8\pi$ as an illustrative example, it is intriguing to observe that under this specific orientation, the mass domains at $\alpha=3\pi/4$ and $7\pi/4$ persist, as demonstrated in Fig.~\ref{fig:neelshift}(c). This noteworthy observation is further bolstered by the real-space energy spectrum and wavefunction distribution, as exemplified in Fig.~\ref{fig:neelshift}(d), thereby providing compelling and convincing evidence for the continued existence of these domains. The conclusion remains valid even when the N\'{e}el vector shifts near the $[11]$ direction, as demonstrated in Figs.~\ref{fig:neelshift}(e) and (f). Hence, our proposal does not require strict alignment of the N\'{e}el vectors along the $[11]$ or $[\bar{1}1]$ directions. Even with deviations from these directions, as long as the N\'{e}el vector is not parallel to the $x$ or $y$ direction, the system still exhibits the distinguishing characteristics of a high-order topological insulator. Notably, when the N\'{e}el vector aligns with the $[11]$ or $[\bar{1}1]$ direction, the system attains the maximum boundary gap, facilitating the detection of corner states.

In the case where the N\'{e}el vector  along the $z$ direction ($\theta=0$), the Hamiltonian in momentum space can be expressed as follows
\eq{
    H(\bfk)=M(\bfk)\sigma_z+A_x\sin k_xs_y\sigma_x-A_y\sin k_ys_x\sigma_x+J_0(\cos k_x-\cos k_y)s_z,\label{eq:hamz}
}
where the last term contributed from altermagnets, and $M(\bfk)=m_0-t_x\cos k_x-t_y\cos k_y$. Similarly, by transforming the Hamiltonian Eq.~\eqref{eq:hamz} into the new coordinates $O-k_\perp k_\parallel$, one can  obtain
\eqa{
    H(k_\parallel,k_\perp)&=M(k_\parallel,k_\perp)\sigma_z+A\cm{k_\parallel\cs{\cos\alpha s_y-\sin\alpha s_x}-k_\perp\cs{\sin\alpha s_y+\cos\alpha s_x}}\sigma_x\\
    &-J_0\cm{\cos2\alpha(k_\perp^2-k_\parallel^2)+2\sin2\alpha k_\parallel k_\perp}s_z,\label{eq:effz}
}
where $M(k_\perp,k_\parallel)=m+1/2t_0(k_\parallel^2+k_\perp^2)$. When the strength of the altermagnetism is small compared with the bulk gap, the Hamiltonian Eq.~\eqref{eq:effz} can be separated into two components, namely $H_0$ and $H_p$
\eqa{
    H_0(k_\parallel,k_\perp)&=M(k_\parallel,k_\perp)\sigma_z-Ak_\perp\cs{\sin\alpha s_y+\cos\alpha s_x}\sigma_x,\\
    H_p(k_\parallel,k_\perp)&=-J_0\cm{\cos2\alpha(k_\perp^2-k_\parallel^2)+2\sin2\alpha k_\parallel k_\perp}s_z+Ak_\parallel\cs{\cos\alpha s_y-\sin\alpha s_x}.
}
After projecting the perturbation term $H_p$ onto the subspace spanned by $\xi_{1,2}$ in the new coordinate system, we obtain the following results
\eqa{
    &\kb{\xi_i|Ak_\parallel(\cos\theta s_y-\sin\theta s_x)\sigma_x|\xi_j}=Ak_\parallel\left(
    \begin{array}{cc}
        1& 0 \\
        0 & -1 \\
    \end{array}
    \right),\\
    &\kb{\xi_i|-J_0\cos(2\alpha)(k_\perp^2-k_\parallel^2)s_z\sigma_0|\xi_j}=0.
}
Therefore, even in the presence of altermagnetism, which breaks time-reversal symmetry, the system still retains edge states without energy gaps when the N\'{e}el vector aligns with the $z$ direction.

\begin{figure}[h]
    \centering
    \includegraphics[scale=0.3]{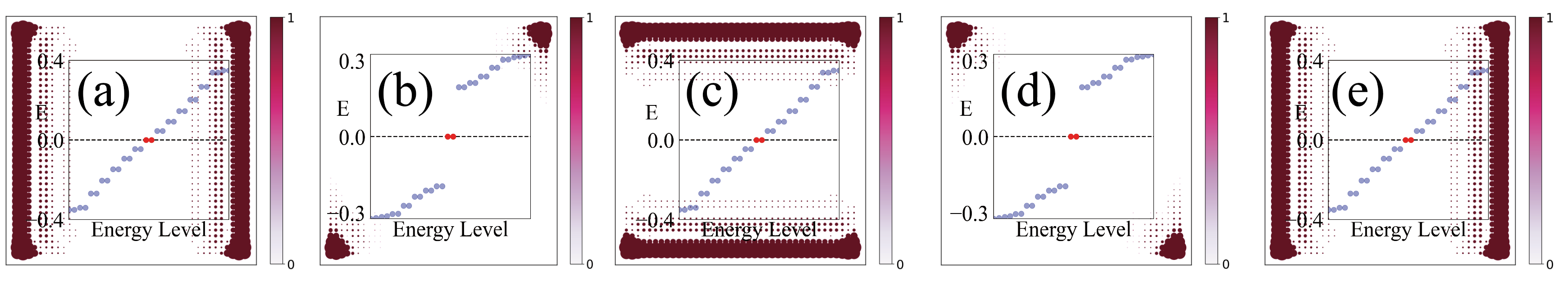}
    \caption{The real-space energy spectrum of the N\'{e}el vector as it undergoes rotation within the plane, along with the corresponding distribution of the zero-energy state wave function. (a) $\varphi=\pi/2$. (b) $\varphi=3\pi/4$. (c) $\varphi=\pi$. (d) $\varphi=7\pi/4$. (e) $\varphi=\pi/2$.    }\label{fig:corner2}
\end{figure}
From a symmetry perspective, it can be deduced that in the absence of altermagnets ($J_0=0$), the Hamiltonian Eq.~\eqref{eq:hamz} exhibits an additional mirror symmetry $\mathcal{M}_z=is_z\sigma_z$, alongside the inherent time-reversal symmetry $\mathcal{T}=is_y\mathcal{K}$. This intriguing observation suggests that the helical edge state is protected by both $\mathcal{T}$ and $\mathcal{M}_z$ symmetries. However, the introduction of an in-plane N\'{e}el component breaks both the time-reversal symmetry $\mathcal{T}$ and the mirror symmetry $\mathcal{M}_z$, resulting in the generation of a gap in the edge states.  For the case of a N\'{e}el component along the out-of-plane direction, it solely breaks the time-reversal symmetry $\mathcal{T}$, while the mirror symmetry $\mathcal{M}_z$ remains. Consequently, the edge state can persist in its gapless nature due to the protection provided by the remaining mirror symmetry $\mathcal{M}_z$.

Based on the edge theory analysis, it has been determined that the characteristics of the edge states are predominantly determined by the in-plane component of the N\'{e}el vector. Consequently, our attention is directed towards investigating the effects of rotating the N\'{e}el vector within the plane.  Upon careful examination, it becomes apparent that as the N\'{e}el vector undergoes rotation within the plane, the corner states experience a gradual transition, eventually transforming into edge states. However, as the rotation continues, the corner states reappear at a different spatial position, as shown in Fig.~\ref{fig:corner2}. This intriguing observation hints at the potential for constructing systems that possess non-Abelian statistics by manipulating the position of the corner states~\cite{wu_double-frequency_2020,wu_non-abelian_2020}.
\section{Mirror-graded winding number when N\'{e}el vector is along $[1\bar{1}]$ direction}\label{sec:mgn}
\begin{figure}[h]
    \centering
    \includegraphics[scale=0.35]{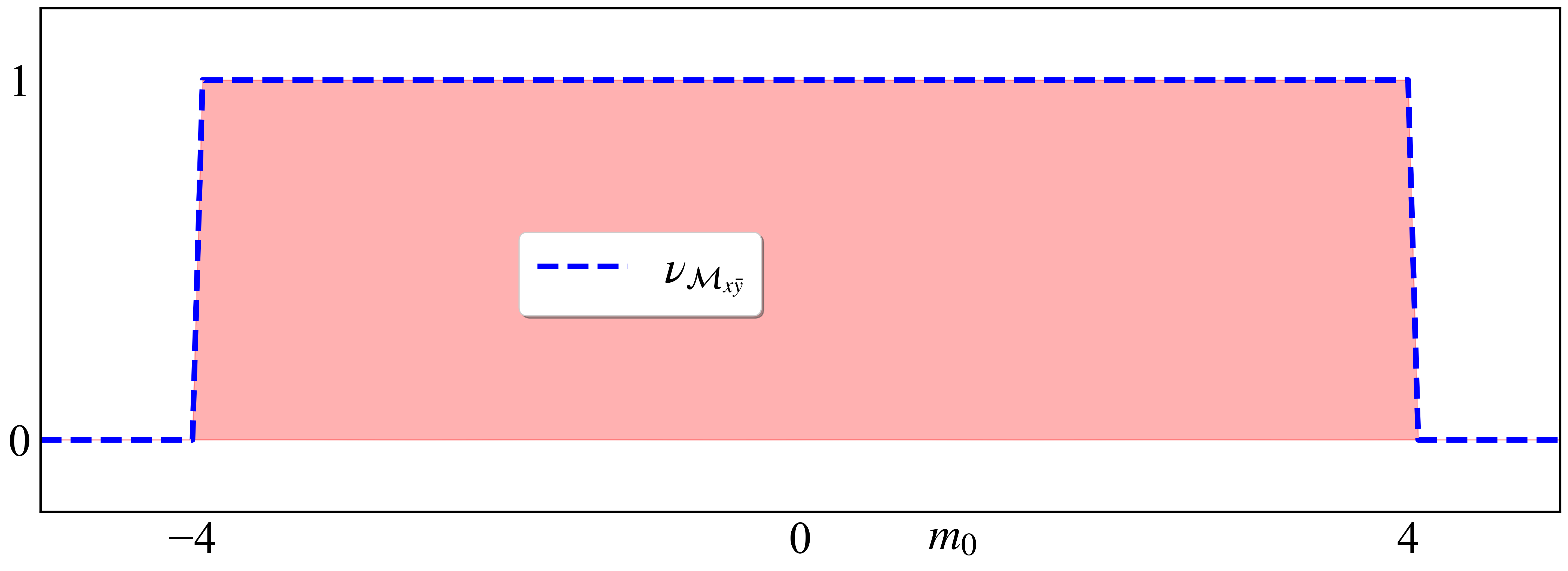}
    \caption{The mirror-graded winding number is plotted as a function of the $m_0$. Common parameters: $\mu=1.0, t_x=t_y=1.0,A=1.0,J_0=0.5$.}\label{fig:mg}
\end{figure}

In this section, we focus on calculating the mirror-graded winding number for the N\'{e}el vector along the $[1\bar{1}]$ direction. For this specific configuration, the Hamiltonian of the system takes the following form
\eqa{
    \mathcal{H}(\bfk)&=\cm{m_0-2t(\cos k_x+\cos k_y)}\sigma_z+2A_0(\sin k_x s_y-\sin k_y s_x)\sigma_x-J(\bfk)s_x\sigma_0+J(\bfk)s_y\sigma_0,\label{eq:q2}
}
where $J(\bfk)=J_0(\cos k_x-\cos k_y)$. The reflection operation $\mathcal{M}_{x\bar{y}}=i\sqrt{2}/2(s_x+s_y)$ maps $(k_x,k_y)$ to $(k_y,k_x)$. Along the reflection symmetry line $k_x=k_y$, the Hamiltonian $\mathcal{H}(\bfk)$ can be decomposed into different mirror subspaces based on the eigenvalues $\pm i$
\eqa{
    H^{\pm i}(k)&=-\cs{m_0-4t\cos k}\eta_z\pm 4A\sin k\eta_x,\label{eq:eff}
}
where $\eta_i$ are Pauli matrice act on mirror eigenvectors.  In each mirror subspace, the Hamiltonian can be expressed in a compact form $H^{\pm i}=\bsf{q}_{\pm i}(k)\cdot\bsf{\sigma }$, the calculated winding numbers are, respectively, $\nu_{+i}=+1$ and $\nu_{-i}=-1$.  Thus, the mirror-graded winding number~\cite{bercioux_topological_2018} can be calculated by $\nu_{\mathcal{M}_{x\bar{y}}}=(\nu_{+i}-\nu_{-i})/2$ with
\eq{
    \nu_{\pm i}=\frac{i}{2\pi}\int_Ldk\cl{\text{Tr}\cm{\bsf{q}_{\pm i}(k)\partial_k\bsf{q}^*_{\pm i}(k)}}.\label{eq:m2}
}
According to Eq.~\eqref{eq:m2}, we plot the variation of the mirror-graded winding number with the $m_0$ in Fig.~\ref{fig:mg}. It is observed that the N\'{e}el vector aligns with $[1\bar{1}]$, when the $m_0\in(-4,4)$, the mirror-graded winding number $\nu_{\mathcal{M}_ {x\bar{y}}}=1$ indicates the presence of a nontrivial topological phase, indicating the existence of TCSs.

\section{The effective Hamiltonian of any boundary for the heterostructure of altermagnets with $g$-wave and $i$-wave as well as $s$-wave spin spliting}\label{edge-gi}

\begin{figure}[h]
    \centering
    \includegraphics[scale=0.33]{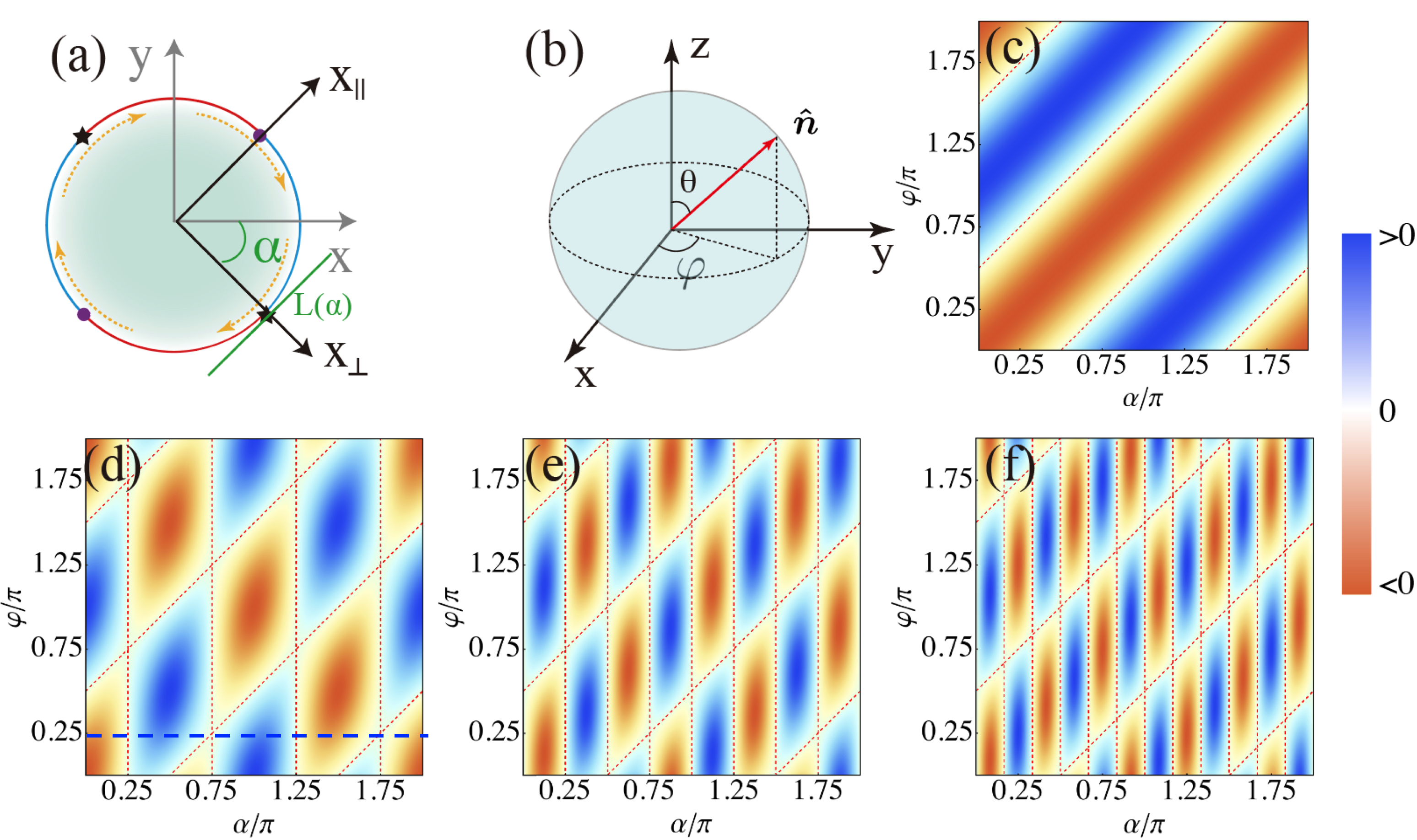}
    \caption{(a) The tangent line L($\alpha$) represents any boundary with the rotated coordinates ($x_{||}$,$x_\bot$) and an angle $\alpha$ to the horizontal axis. (b) The unit vector $\bsf{\hat{n}}$ of the N\'{e}el vector with the polar angle $\theta$ and azimuthal $\varphi$ angles in the spherical coordinates.  (c)-(f) Variation of the Dirac mass $\mathcal{L}(\alpha,\theta=\pi/2,\varphi)$ on any boundary  $L(\alpha)$ with the azimuth of the N\'{e}el vector (or magnetization direction). (c) Zeeman filed ($s$-wave spin splitting). (d) $d$-wave spin splitting. (e) $g$-wave spin splitting. (f) $i$-wave spin splitting. The red dashed line denotes a zero mass term.  A domain wall emerges and localizes a corner state when the sign of the masses on the left/right of the red dashed lines is opposite.}\label{fig:igwave}
\end{figure}

For the heterostructures of altermagnets with the spin splitting of other symmetries, including $g$-wave, and $i$-wave~\cite{smejkal_beyond_2022} which corresponding orbital quantum number ${\it l}=4,6$, respectively,  the boundary Hamiltonian can be obtained by rotating the coordinate system as done for the $d$-wave in the text.  We derive the boundary Hamiltonian of the spin splitting with a non-zero orbital quantum number for any boundary $L(\alpha)$ and any N\'{e}el vector with polar ($\theta$) and azimuthal ($\varphi$) angles
\eqa{
    H_{\rm eff}(x_\perp,k_\parallel)=Ak_\parallel\eta_z+\mathcal{L}(\alpha,\theta,\varphi)\eta_y,
}
where
\eq{
    \left\{
    \begin{array}{c}
        {\it l}=2:\quad   \mathcal{L}(\alpha,\theta,\varphi)\sim \cos(2\alpha)\sin(\theta)\cos\cs{\varphi-\alpha}\quad d-\text{wave},\\
        {\it l}=4:\quad     \mathcal{L}(\alpha,\theta,\varphi)\sim \sin(4\alpha)\sin(\theta)\cos\cs{\varphi-\alpha}\quad g-\text{wave},\\
        {\it l}=6:\quad     \mathcal{L}(\alpha,\theta,\varphi)\sim\sin(6\alpha)\sin(\theta)\cos\cs{\varphi-\alpha}\quad i-\text{wave},\\
    \end{array}
    \right.\label{eq:igwave}
}
and $\eta_i$ are Pauli matrices. According Eq.~\eqref{eq:igwave} one can obtain that for a N\'{e}el vector confined to the plane ($\theta=\pi/2$), the Dirac mass is equal to zero when $\varphi-\alpha=(2n+1)\pi/2,n\in\mathbb{Z}$, where boundary $L(\alpha)$ is perpendicular to the N\'{e}el vector.  This situation is independent of the form of spin splitting.  Roughly speaking, the Dirac mass can be divided into two parts: one that depends on the azimuthal angle $\mathcal{L}^0(\alpha,\varphi)$ and one that does not $\mathcal{L}^0(\alpha,\theta)$.  The $\varphi$-independent part contributes a factor that depends on the boundary  $L(\alpha)$, which is consistent with the variation of the spin splitting magnitude with the direction of the momentum~\cite{smejkal_beyond_2022}.
For spin splitting with higher orbital quantum number $(l=4,6)$, there will be more locations where the splitting is equal to zero, thus contributing more corner states.   As shown in Figs.~\ref{fig:igwave}(d)-(e), we plot the density map of the Dirac mass $\mathcal{L}(\alpha,\pi/2,\varphi)$ with respect to the boundary  $L(\alpha)$ and the N\'{e}el vector azimuthal angle $\varphi$. The red dashed line represents where the Dirac mass is equal to zero.  Given an azimuthal angle $\varphi_0$, if the solutions to $\mathcal{L}^0(\alpha,\varphi_0)=0$ and $\mathcal{L}^0(\alpha,\theta=\pi/2)=0$ coincide at a specific value of $\alpha$ (denoted by $\alpha_0$ here), then the Dirac masses on both sides of the boundary at $\alpha_0$ will have the same sign. Consequently, no domain wall will form at this point, as illustrated by the blue dashed line in Fig.~\ref{fig:igwave}(d). As the azimuth deviates from the special value $\varphi_0$, a new pair of domain walls will be created at $\alpha_0$, i.e. the number of corner states will increase. Consequently, as the azimuthal angle $\varphi$ varies, domain walls will undergo continuous creation, movement, and annihilation. This dynamic behavior is mirrored by the corner states.

The band splitting produced by the Zeeman field can be regarded as an isotropic $s$-wave with orbital quantum number $(l=0)$, where the boundary Hamiltonian has the same form as that of the above altermagnets with the Dirac mass as follows
\eq{
    \mathcal{L}(\alpha,\theta,\varphi)\sim\sin(\theta)\cos\cs{\varphi-\alpha}\quad s-\text{wave}.\label{eq:swave}
}
According to Eq.~\eqref{eq:swave}, the only corner states occur when the boundary $L(\alpha)$ is perpendicular to the N\'{e}el vector. As the N\'{e}el vector orientation changes, the corner states will correspondingly move along the boundary. Hence, our study encompasses the scenario where the Zeeman field-induced spin splitting is treated as an $s$-wave with orbital quantum number $l=0$. This specific case falls within the broader framework of our investigation.

\section{First-principle calculations methods}\label{sec:dft}
\begin{figure}[h]
    \centering
    \includegraphics[scale=0.35]{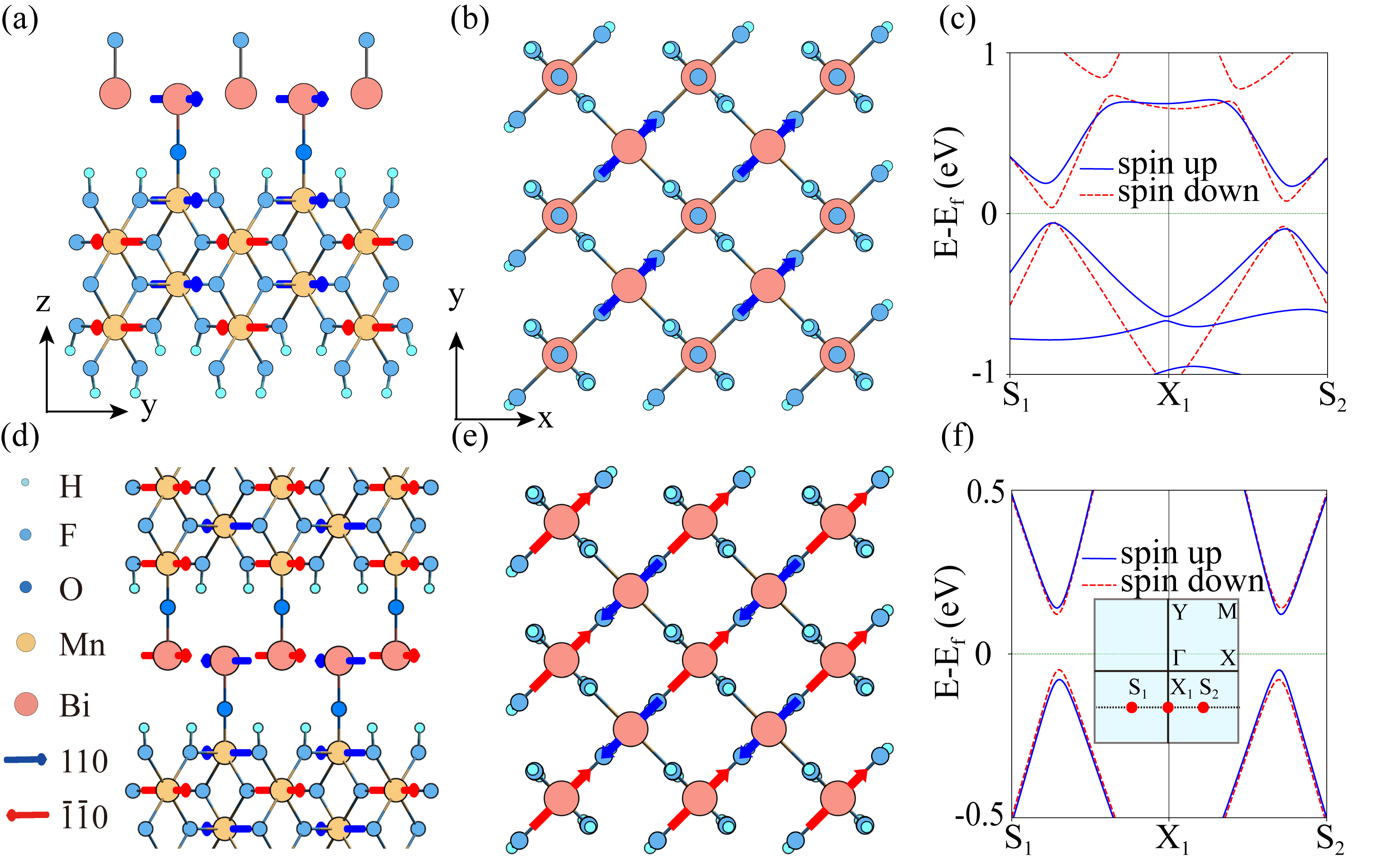}
    \caption{(a) Side view and (b) top view of the structure of Bi/MnF$_2$. (c) Band strucure of the Bi/MnF$_2$ heterstructure. (d) Side view and (e) top view of the structure of MnF$_2$/Bi/MnF$_2$. (f) Band strucure of the MnF$_2$/Bi/MnF$_2$ sandwich structure. The insert shows the 2D Brillouin zone with high-symmetric points and lines.
    }\label{fig:S1}
\end{figure}

We perform first-principle calculations using the projector augmented wave method~\cite{blochl_projector_1994} implemented in the Vienna \textit{ab initio} simulation package (VASP)~\cite{kresse_ab_1994,kresse_efficient_1996}.
Generalized gradient approximation (GGA) of the Perdew-Burke-Ernzerhof (PBE) functional ~\cite{perdewGeneralizedGradientApproximation1996} is selected as the exchange-correlation potential. The plane-wave cutoff energy is set to 550 eV, and the Monkhorst-Pack k-point mesh~\cite{monkhorst_special_1976}
of size 9 $\times$ 9 $\times$ 1 is used in Brillouin zone for self-consistents. The optimized lattice constant of MnF${}_2$/Bi/MnF${}_2$ is 4.95 \AA, and the buckled of the Bi layer is 0.2 \AA.
Considering the correlation effect of $d$ electrons, we use the DFT+U method~\cite{anisimov_band_1991,dudarev_electron-energy-loss_1998} in the calculation and choose $U =$ 2 eV for Mn-3$d$ orbitals.

\section{A sandwich structure of a 2D TI and altermagnets}\label{sec:sand}
Initially, a heterostructure was employed by placing bismuthene on the surface of an altermagnet MnF$_2$, as shown in Figs.~\ref{fig:S1}(a) and (b). It can be seen that only these Bi atoms directly bonded to Mn through O atoms exhibit magnetization, leading to a nonzero magnetization and a ferromagnetic band structure, as shown in Fig.~\ref{fig:S1}(c). To induce altermagnetism,  a sandwich structure with $S_{4}$ symmetry was adopted, with Mn atoms flanking the Bi layer exhibiting antiparallel spins, ensuring the zero net magnetization, as depicted in Figs.~\ref{fig:S1}(d) and (e). In this structure, Mn atoms induced opposite magnetic orders on the two Bi atoms within a unit cell, which were connected via $[C_2][S_{4}]$ symmetry, giving rise to a $d$-wave altermagnetism. Here $C_2$ denotes the $180^\circ$ rotation of spin directions in spin space and $S_4$ connects the sublattices with opposite spins in real space. The altermagnetic band structure is shown in Fig.~\ref{fig:S1}(f).

\section{\textit{Ab initio} tight-binding model for the sandwich structure}\label{sec:tb}

\begin{figure}[h]
    \centering
    \includegraphics[scale=0.5]{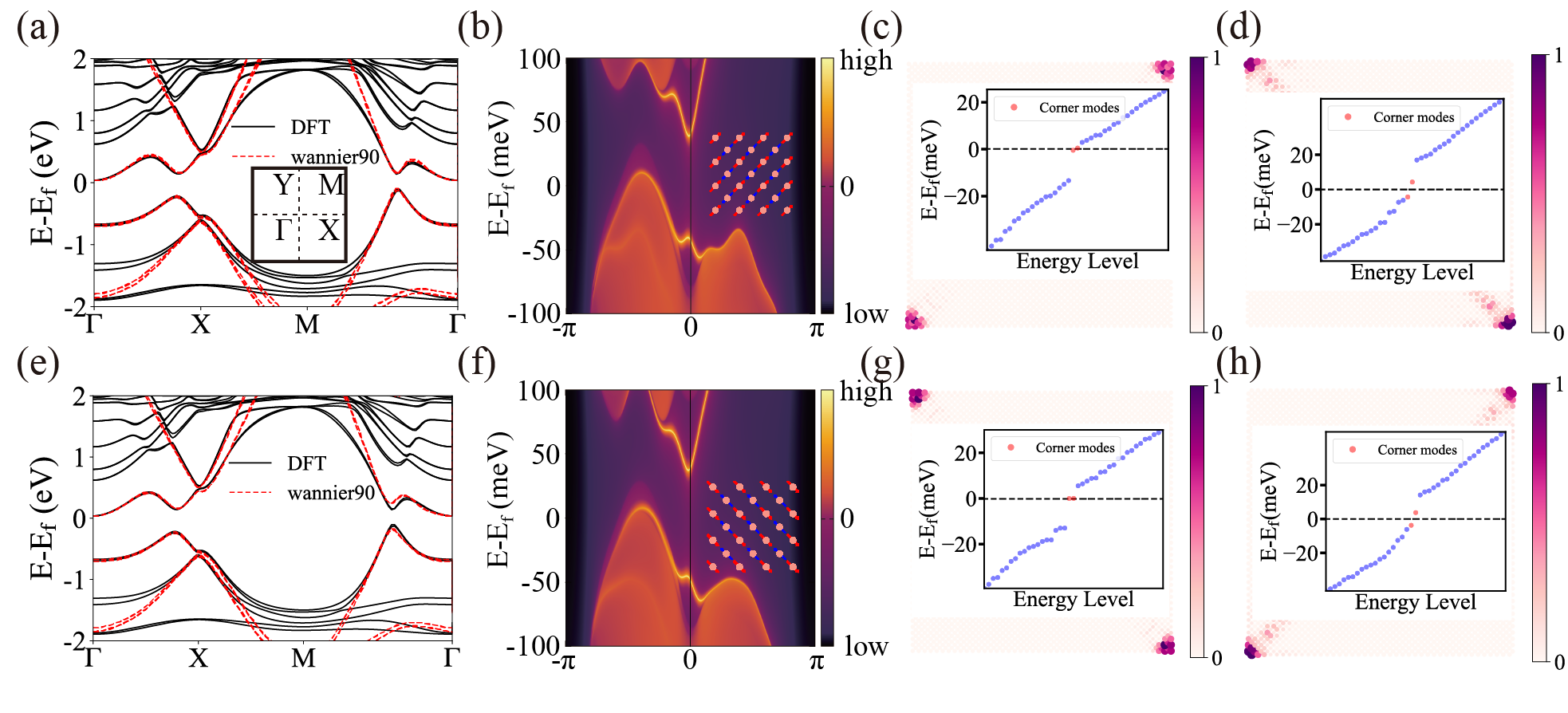}
    \caption{The first row shows the case with the N\'{e}el vector along the diagonal direction. (a) The band structures were calculated by DFT (black solid line) and WANNIER90 (red dashed line). The insert shows the 2D Brillouin zone with high-symmetric points and lines. (b) The edge spectrum of MnF$_2$/Bi/MnF$_2$ sandwich structure. Inset: The small arrows depict the direction of the N\'{e}el vector. (c) Spectrum for a finite-size square sample. Inset: Two in-gap topological corner states emerge. The real spatial distribution of their wave function is plotted. (d) The spectrum of a finite-size square sample that exposes different atoms compared to (c). (e)-(h) Same as (a)-(d) except the N\'{e}el vector along the off-diagonal direction.}\label{fig:S2}
\end{figure}
Based on the DFT calculation of MnF$_2$/Bi/MnF$_2$ sandwich structure, an \textit{ab initio} tight-binding (TB) model of the Bi-6$p_x$ and Bi-6$p_y$ orbitals was constructed here using the WANNIER90 program~\cite{marzari_maximally_1997-1,souza_maximally_2001}. Using this TB model, the band structures, edge states, and the real space spectrums with different orientations of the N\'{e}el vector were calculated. Figures.~\ref{fig:S2}(a), (b), (c) and (d) present the results for the N\'{e}el vector along the diagonal direction, while Figs.~\ref{fig:S2} (e), (f), (g), and (h) for the off-diagonal direction. It can be observed that the altermagnetism in the Bi layer resulted in a $30$~meV energy gap in the edge states [see Fig.~\ref{fig:S2}(b) and (e)].  We calculated the energy spectrum of the finite size structure, and the wavefunction of in-gap states mainly distributed at corners as shown in Fig.~\ref{fig:S2}(c). Given that there are multiple atoms in a unit cell when different atoms are exposed at the boundary, the wave function distribution of in-gap states will change, as shown in Fig.~\ref{fig:S2}(d). Furthermore, when altering the direction of the N\'{e}el vector into the off-diagonal direction, the gapped edge states remained[see Fig.~\ref{fig:S2}(f)], and the spatial distribution of the corner states was regulated as shown in Figs.~\ref{fig:S2}(g) and (h). In conclusion, manipulating the direction of the N\'{e}el vectors can serve as an efficient means to regulate the behavior of the corner states.

\section{A minimal tight-binding model  for the sandwich structure}\label{sec:tb_model}

\begin{figure}[h]
    \centering
    \includegraphics[scale=0.35]{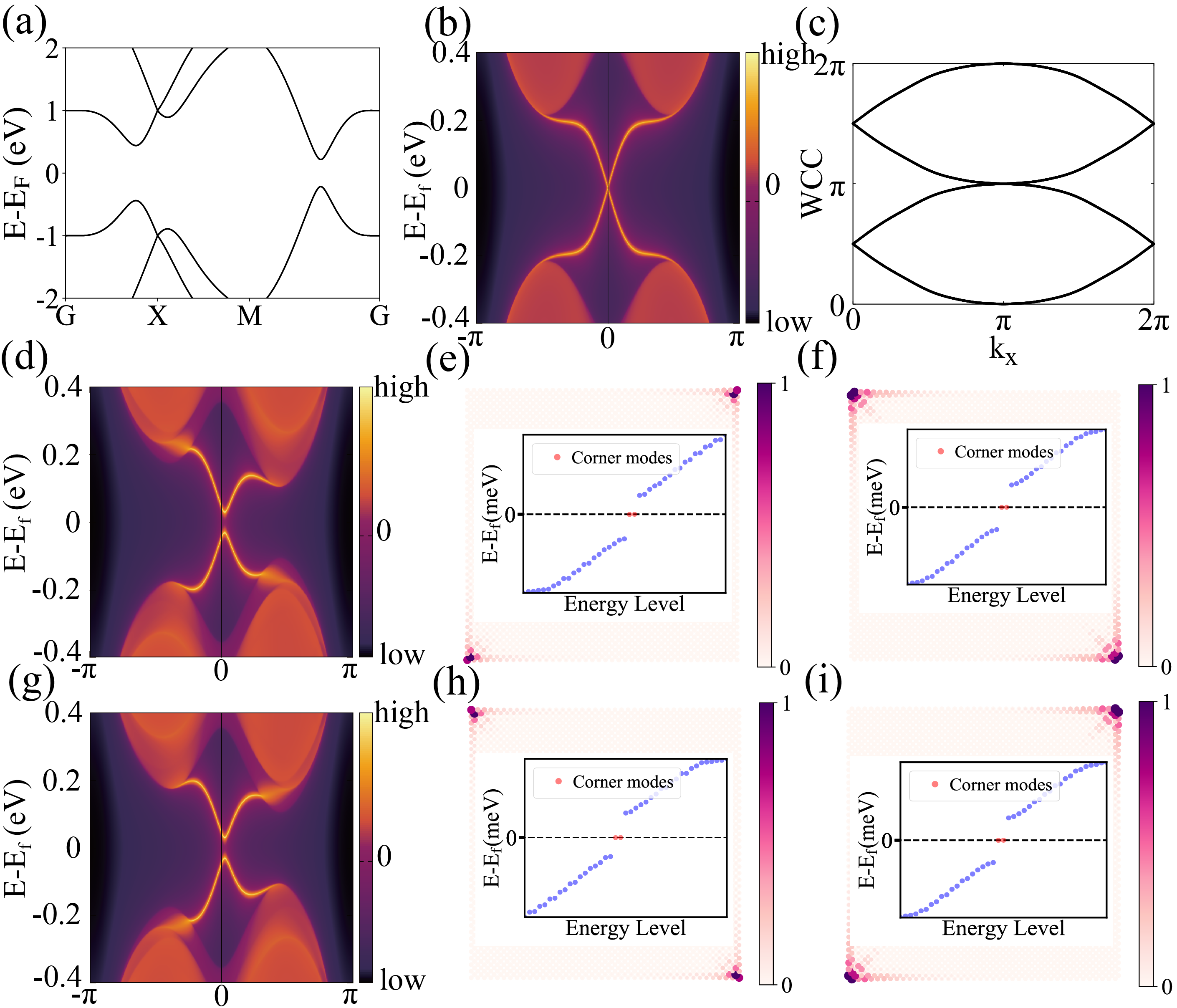}
    \caption{(a)-(c) show the results of tight-binding model under the parameters $\lambda_\text{SOC}=t=1$ eV, and $r_1=r_2=0.3$ eV. (a) The band structure was calculated. (b) The edge spectrum of the model with gapless edge state. (c) The Wilson loop of the TI state is calculated by the model. (d)-(f) and (g)-(i) respectively show the results of the N\'{e}el vector along the diagonal and off-diagonal lines under $m=-J_0=0.05$ eV after adding the altermagnetism term. (d) The gapped edge spectrum of the model with the diagonal N\'{e}el vector. (e) Spectrum for a finite-size square sample and Inset: Two in-gap topological corner states emerge. The real spatial distribution of their wave function is plotted. (f) Same as (e) except changed the boundary atoms. (g)-(i) The results same as (d)-(f) except the N\'{e}el vector is along off-diagonal direction.}\label{fig:S3}
\end{figure}
In this section, a minimal tight-binding model is constructed to study the relationship between the N\'{e}el vector and the corner state. In the basis of
$\frac{1}{\sqrt{2}}\{\ks{p_x^A\uparrow}+\ks{p_y^A\uparrow},\ks{p_x^A\uparrow}-\ks{p_y^A\uparrow},\ks{p_x^B\uparrow}+\ks{p_y^B\uparrow},\ks{p_x^B\uparrow}-\ks{p_y^B\uparrow},\ks{p_x^A\downarrow}+\ks{p_y^A\downarrow},\ks{p_x^A\downarrow}-\ks{p_y^A\downarrow},\ks{p_x^B\downarrow}+\ks{p_y^B\downarrow},\ks{p_x^B\downarrow}-\ks{p_y^B\downarrow}\}$($A$ and $B$ label the two atoms in the unit cell, and $\uparrow,\downarrow$ labels spin-up and spin-down), the Hamiltonian has the following form
\eqa{
    H(\bfk)&=-\lambda_\text{SOC}\sigma_y s_z+4 t\tau_x\left(\cos\frac{k_x}{2}\cos\frac{k_y}{2}-\sigma_z\sin\frac{k_x}{2}\sin\frac{k_y}{2}\right)+H_\text{Rashba}(\bfk),\\
    H_\text{Rashba}(\bfk)&=r_1\tau_z\sigma_z\left( s_x\sin k_x-s_y\sin k_y\right)+r_2\tau_z\left(s_y\sin k_x-s_x\sin k_y\right),\label{Bi2F2}
}
where $\tau$, $\sigma$, and $s$ label sublattices, orbital and spin degrees of freedom, respectively.  The energy band along the high symmetry path is shown in Fig.~\ref{fig:S3}(a), which is consistent with the DFT result near the Fermi surface. In Figs.~\ref{fig:S3}(b) and (c), the edge states of the semi-infinite system and the Wilson loop of bulk are calculated, confirming that the system is a non-trivial topological insulator.

To simulate altermagnetism in the MnF$_2$/Bi/MnF$_2$ sandwich structure, the following form is considered
\eq{
    H_\text{AM}=m\tau_z\sigma_z+J_0\tau_z\bm s\cdot\hat{\bm n},\label{AFM}
}
where the first term breaks inversion symmetry while preserving $S_4$ symmetry that couples the two sublattices, in contrast, the last term represents the opposing spin between the two sublattices. The energy spectrum of the semi-infinite system when the N\'{e}el vector is along the $[11]$ and $[1\bar{1}]$ directions are shown in Figs.~\ref{fig:S3}(d) and (g), respectively. The introduction of altermagnets breaks the time-reversal symmetry and leads to a gap of around $30$ meV. To study corner states, structures of finite size are considered and their energy spectrum is calculated as shown in Figs.~\ref{fig:S3}(e,f) and Figs.~\ref{fig:S3}(h,i). It can be seen that there are two in-gap states in the energy spectrum and their wave functions are located at the corners of the system.  Similar to the DFT calculation, we also analyze the boundaries in Figs.~\ref{fig:S3}(f) and (i) to account for the presence of different atoms. This analysis reveals that, for a fixed N\'{e}el vector, the boundary conditions affect the wave function distribution of the in-gap states. This finding is consistent with the results obtained from the DFT calculations.
    \section{symmetry and topological classification}\label{sec:sw}
    \begin{figure}[t]
        \centering
        \includegraphics[scale=0.25]{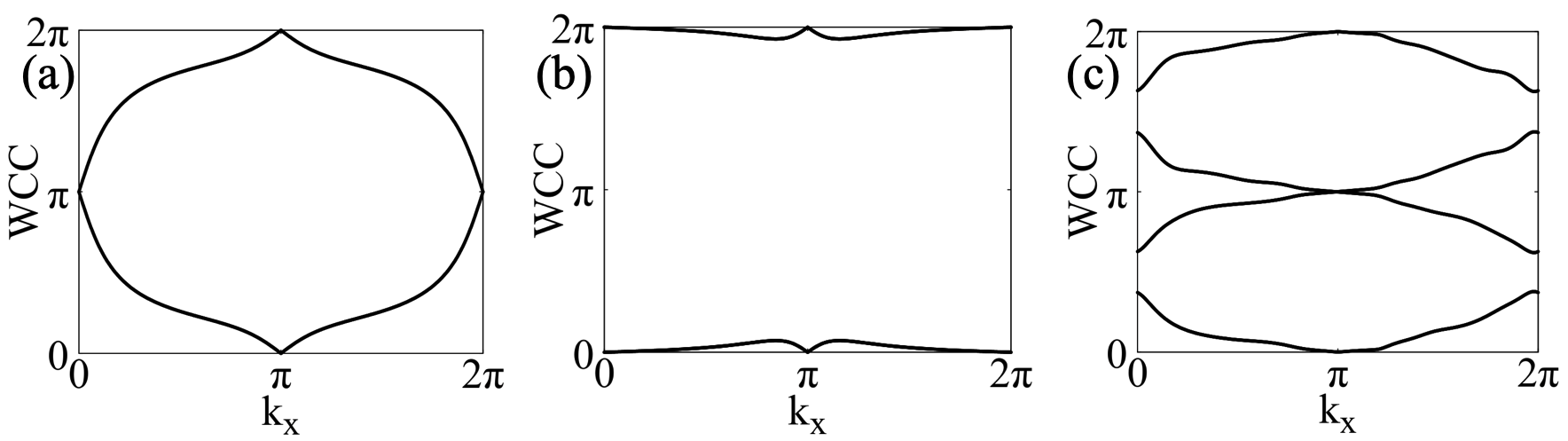}
        \caption{Wilson loop calculations for higher-order topology. (a) Higher-order topological insulator phase: The Wilson loop calculated using our model intersects the $\theta=\pi$  an odd number of times.  (b) Trivial phase:  The Wilson loop in the trivial phase intersects the $\theta=\pi$ line an even number of times. (c) Wilson loop calculations for MnF$_2$/Bi/MnF$_2$ sandwich structure, and intersects the $\theta=\pi$  an odd number of times.}\label{fig:s11}
    \end{figure}
    In this section, we investigate the topological classification of the system and compute the topological invariant using the Wilson loop method.   When the N\'{e}el vector is confined to the $x-y$ plane, the Hamiltonian~\eqref{eq:q2} exhibits $C_{2z}T=s_x\mathcal{K}$ symmetry with $\mathcal{K}$ representing complex conjugate and belongs to the Stiefel-Whitney (SW) classification~\cite{bouhon_wilson_2019,ahn_symmetry_2019}. The topology can be characterized by the second SW invariant $w_2$. To demonstrate this more clearly, we perform a unitary transformation on $H(\bfk)$ using the operator $U=\exp(i\pi/4 s_x)$, and obtain its real form $H^\prime(\bfk)=UH(\bfk)U^\dagger$ read as
    \eqa{
        H^\prime(\bfk)&=(m_0-2t_x\cos k_x-2t_y\cos k_y)\sigma_z+A_x\sin k_xs_z\sigma_x-A_y\sin k_ys_x\sigma_x\\
        &+2J_0(\cos k_x-\cos k_y)s_x+2J_0(\cos k_x-\cos k_y)s_z.\label{eq:realham}
    }
    The transformed Hamiltonian $H^\prime(\bfk)$  preserves $C_{2z}T=\mathcal{K}$ symmetry, and its components are all real numbers.  The Wilson loop can serve as a diagnostic tool for calculating the second  SW number $w_2$. The number of times the Wilson loop intersects with $\theta=\pi$ dictates the value of $w_2$:  an odd number of intersects signifies $w_2=1$, while an even number signifies $w_2=0$. A non-zero $w_2(w_2=1)$ signifies a non-trivial SW insulator, characterized by the presence of corner
    states.

    For the Hamiltonian given by Eq.~\eqref{eq:realham}, the Wilson loop along the $k_x$ direction was calculated for both the higher-order topological phase and the trivial phase, as shown in Figs.~\ref{fig:s11}(a) and (b). It can be observed that in the higher-order topological phase, the Wilson loop intersects $\theta=\pi$ an odd number of times, corresponding to $w_2=1$, while in the trivial phase $w_2=0$. The Wilson loop for our devised MnF$_2$/Bi/MnF$_2$ sandwich structure was also calculated, confirming $w_2=1$, as shown in Fig.~\ref{fig:s11}(c).

\end{document}